\documentclass[12pt,letterpaper]{article}
\usepackage[a4paper, total={7in, 10in}]{geometry}

\usepackage{graphicx}
\usepackage{helvet}
\usepackage{authblk}
\usepackage{hyperref}
\usepackage{amsmath}
\usepackage{amssymb}
\usepackage{siunitx}
\usepackage{booktabs}
\usepackage{orcidlink}
\usepackage[super,comma,sort&compress]{natbib}

\DeclareSIUnit{\Ah}{Ah}
\newcommand{\dvdsoc}{$\mathrm{d}V/\mathrm{d}\mathrm{SOC}$}

\makeatletter
\renewcommand{\maketitle}{\bgroup\setlength{\parindent}{0pt}
\begin{flushleft}
  \textbf{\@title}

  \@author
\end{flushleft}\egroup}
\makeatother

\title{Estimating the Health and State of Charge of Each Cell
in a Second-Life Battery System from Field Data}
\date{}

\author[1,*]{Martín Cornejo}
\author[2]{Julian Meyer-Schwickerath}
\author[1]{Juan Victor Sandalinas}
\author[1]{Andreas Jossen}

\affil[1]{Technical University of Munich, TUM School of Engineering and Design,
Department of Energy and Process Engineering, Chair of Electrical Energy Storage Technology,
Arcisstr. 21, 80333 Munich, Germany}
\affil[2]{STABL Energy GmbH, Baierbrunner Str. 30, 81379 Munich, Germany}
\affil[*]{Correspondence: martin.cornejo@tum.de}

\begin{document}

\maketitle

\section*{SUMMARY}

Effective use of battery storage depends on reliable estimation of its state of health (SOH) and state of charge (SOC).
Model-based state estimation requires the open-circuit voltage (OCV) curve,
which is typically unknown for second-life batteries.
We present a framework that jointly estimates the states and parameters of an equivalent circuit model
solely from field operation data, using Gaussian process regression to reconstruct the OCV curve.
Applied to a real second-life battery system of 27 modules and 324 cells, it reveals SOH heterogeneity,
a systematic SOC imbalance, and two faulty cells, all validated against a reference measurement.
We aggregate the cell SOH and SOC to module level
and benchmark them against a lumped-module model fitted without the individual cell voltages.
The lumped-module model follows the average behavior and cannot capture the limiting cells,
overestimating SOH by up to \SI{31}{\percent} and SOC by up to \SI{23}{\percent}.

\section*{KEYWORDS}

lithium-ion batteries, 
second-life batteries, 
field data, 
data-driven,  
equivalent circuit model,
Gaussian processes, 
extended Kalman filter, 
state of health,
state of charge, 
open-circuit voltage

\section*{INTRODUCTION}

Decarbonizing the power grid requires flexible and cost-effective energy storage at scale.
At the same time, the rapid electrification of transport is generating a growing supply
of retired battery packs with significant residual capacity.
Second-life batteries have long been proposed as a sustainable solution for stationary storage,
yet their widespread adoption remains limited by challenges related to cost competitiveness, safety, and reliability
\cite{zhu_endoflife_2021, borner_challenges_2022}.
Among these, a central difficulty is that cells entering a second-life system carry heterogeneous degradation states.
Differing usage histories, combined with intra-pack thermal gradients and manufacturing tolerances during first life,
lead to divergent degradation trajectories
\cite{mowri_assessing_2024, naylor_marlow_degradation_2024, akhtarizavareh_heterogeneous_2026}.

This heterogeneity adds complexity to the monitoring and operation of second-life storage systems.
Weaker cells reach their operational limits sooner and constrain the full pack, reducing overall system capability
\cite{ruther_battery_2025, feng_propagation_2019}.
Beyond capacity and resistance, cells also drift apart in their state of charge (SOC),
and this SOC imbalance further reduces the usable capacity of the pack.
Modules and packs are commonly modeled as a single large cell,
which is adequate as long as the cells behave alike \cite{su_stateofhealth_2024}.
For second-life batteries, however, the system-level SOC and state of health (SOH)
may no longer be adequately represented by an average cell model.
Ignoring this cell-to-cell variation may lead either to unexpected capacity shortfalls or, 
if the system is operated conservatively, to underutilization of available energy \cite{dubarry_battery_2019}.
In practice, neither the prior usage history nor the current degradation state of individual cells is typically known 
at the time of repurposing, and even a known state would not remain valid, 
as the cells continue to diverge in their second life.

Battery management systems (BMS) continuously collect voltage, current, and temperature during operation,
providing limited and noisy measurements from which SOC and SOH must be estimated.
A range of data-driven and model-based methods has been proposed for this purpose.
Purely data-driven approaches, particularly deep learning, 
have attracted significant attention but rely on large labeled datasets and offer limited physical interpretability, 
restricting their use outside controlled laboratory conditions \cite{yao_smarter_2025}.
Model-based approaches pair a cell model with a state estimator
that fuses model predictions with incoming measurements.
Kalman filter variants are the most established class, with benchmark studies confirming their accuracy
under proper tuning \cite{hossain_kalman_2022, campestrini_comparative_2016}.

Estimator accuracy depends directly on model quality: 
a model that no longer reflects the cell's true behavior produces biased state estimates regardless of filter design.
The equivalent circuit model (ECM) is the standard choice for battery state estimation
due to its physical interpretability and computational efficiency.
It models the terminal voltage as the sum of the open-circuit voltage (OCV), an internal resistance drop,
and one or more resistance-capacitance (RC) pairs to describe transient polarization.
This captures the main electrical behavior of a cell with a small number of identifiable parameters.
The parameters, however, drift with degradation.
Capacity fades and resistance grows, so a fixed parameter set produces growing estimation errors.
A dual filter addresses this by augmenting the state vector with parameters and updating them alongside SOC
\cite{plett_extended_2004}.
A further complication is that ECM parameters are functions of the operating conditions rather than scalar constants.
Internal resistance depends on SOC and temperature, and the OCV is a nonlinear function of SOC.
Conventional parametric forms can only approximate these dependencies \cite{peng_comprehensive_2023}.
This rigidity can be addressed by embedding data-driven components within the ECM structure,
replacing selected parametric terms with neural networks \cite{kuzhiyil_neural_2024}
or non-parametric Gaussian processes (GPs) \cite{aitio_learning_2025, schaeffer_health_2024, zhou_learning_2025}.
Yet Zhou et al.\ expose a limitation: 
in an aging-aware GP-ECM designed to estimate capacity and resistance from cycling data, 
aging-induced changes in the OCV curve are absorbed into the learned resistance, 
biasing its estimate \cite{zhou_learning_2025}.
A beginning-of-life OCV curve cannot be relied upon as the cell ages
and must be adapted alongside the other parameters.

Recovering the OCV from operational data has therefore attracted growing interest.
The most straightforward approaches fit OCV points to a parametric form such as a polynomial
\cite{chen_novel_2019, yang_open_2026, wang_state_2024},
but can only coarsely approximate the detailed shape of the OCV curve.
More accurate methods typically require specific operating conditions such as rest periods or low-current phases 
\cite{wang_novel_2023, hofmann_deltaq_2024, figgener_degradation_2024}, 
and additionally depend on chemistry-specific calibration \cite{wang_novel_2023}
or prior laboratory half-cell measurements \cite{hofmann_deltaq_2024},
neither of which is generally available for second-life cells.
In prior work, we treated the OCV curve and the SOC-dependent resistance as Gaussian processes
identified jointly from arbitrary cycling data, requiring no pre-measured reference curve
and achieving below \SI{10}{\milli\volt} terminal voltage RMSE on NMC cells
across SOH \SIrange{72}{100}{\percent} \cite{cornejo_datadriven_2025}.
However, the approach was only validated on individual cells under laboratory conditions.

In this work, we present a recursive GP-ECM framework that jointly estimates SOC and ECM parameters
while reconstructing the OCV curve directly from field operation data, requiring no prior cell characterization.
To capture how the differences between cells constrain the system capability,
we describe every cell with its own model and aggregate the cell models to module and system level.
We demonstrate the framework on a real second-life battery energy storage system (BESS) of 27 modules and 324 cells.
From a \SI{12.5}{\hour} measurement window, we estimate the SOH heterogeneity and the SOC imbalance across all cells,
and validate these estimates, together with the reconstructed OCV curves, against a dedicated reference measurement.
As a benchmark, we compare the cell-based system model against the simpler alternative of a lumped-module model,
a single model per module fitted without the individual cell voltages.
We evaluate both on voltage prediction accuracy and on the estimated module SOH and SOC.

\section*{RESULTS}
\subsection*{Battery system and dataset}

The system under study is a second-life BESS research prototype developed and operated by STABL Energy GmbH.
It comprises 27 battery modules arranged in three phases and delivers a peak power of \SI{67.5}{\kilo\watt}.
The system employs a modular multilevel converter (MMC) topology, in which each module incorporates its own H-bridge 
power electronics and is switched individually into and out of the series string to synthesize the AC output.
The switching order across modules can be set freely and is used here for SOC balancing: 
modules with a higher SOC, as estimated by the onboard BMS, are assigned a larger share of the load.
Each module consequently alternates between periods of full load, partial load,
and rest even during continuous system power delivery.
In conventional series-connected systems, by contrast, 
all modules share a common current and the weakest module limits usable capacity.
The MMC's flexibility therefore makes it well suited to second-life battery systems,
where remaining capacity and resistance vary across cells \cite{li_soh_2018, rivera_state_2026}.

Each of the 27 modules contains 12 \textit{logical} cells in series, for a total of 324 logical cells across the system.
Each logical cell consists of two LEV50N \textit{physical} cells connected in parallel (2P configuration),
sharing a common terminal voltage and forming a single observable unit with a nominal capacity of \SI{100}{\Ah}.
Throughout this paper, ``cell'' refers to a logical cell unless otherwise specified.
The LEV50N is a prismatic LMO/graphite cell rated at \SI{50}{\Ah}, introduced by GS Yuasa in 2012 
\cite{ueki_development_2012} and widely deployed in electric vehicles such as the Mitsubishi i-MiEV
and Peugeot iOn during the first half of the 2010s \cite{fong_study_2020, guenther_second_2024}.
The cells were sourced without any record of prior usage, capacity, or degradation state,
so their condition was unknown at the time of integration.

\begin{figure}
    \centering
    \includegraphics[width=\textwidth]{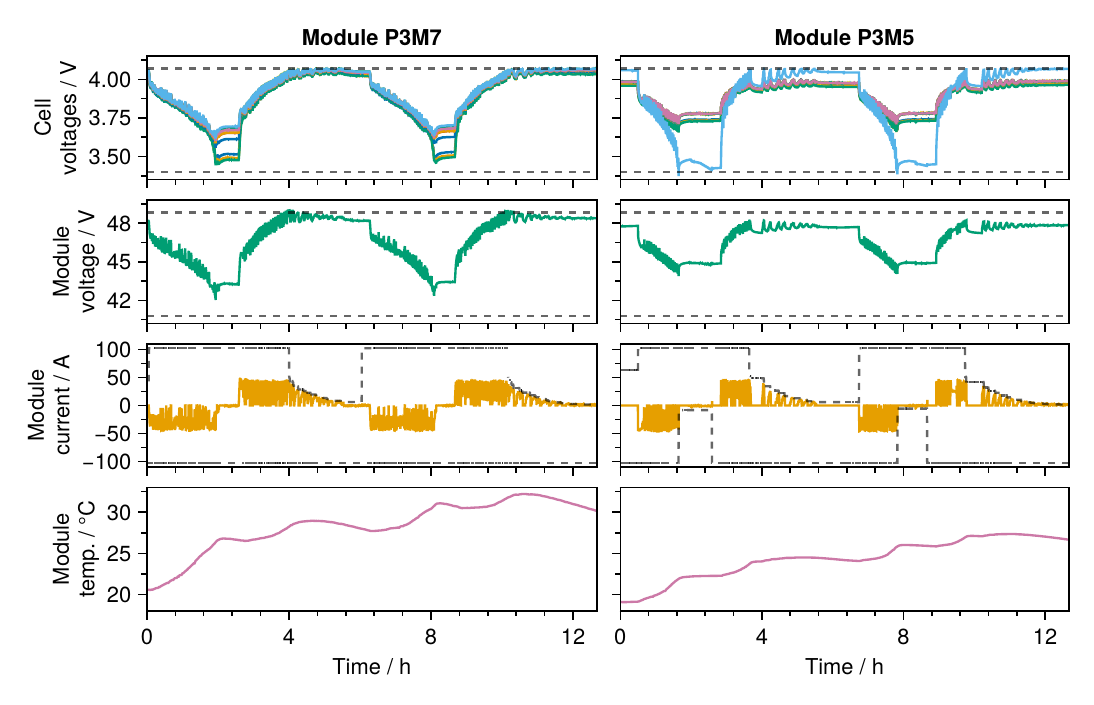}
    \caption{Operation of two example modules over the measurement window:
    P3M7, whose cell voltages spread apart only near the end of discharge,
    and P3M5, which contains the divergent cell P3M5C12 (light blue).
    From top to bottom: the twelve cell voltages, module voltage, module current, and module temperature.
    Dashed lines mark the cell and module voltage limits, and the current limits set by derating.
    In P3M5, the divergent cell reaches the voltage limits earlier than the rest,
    which triggers derating earlier and reduces the module's charge throughput.}
    \label{fig:dataset}
\end{figure}

For this study, the system was operated over a \SI{12.5}{\hour} window comprising two charge\slash{}discharge cycles
at an approximately constant power of \SI{35}{\kilo\watt}, corresponding to a phase current of about \SI{50}{\ampere}.
Although the phase current was held approximately constant, individual module currents vary dynamically as the MMC 
continuously adjusts its switching sequence, producing a load profile representative of field operation
(Figure~\ref{fig:dataset}).
Cycles were bounded by cell-level voltage limits of \SI{3.4}{\volt} and \SI{4.07}{\volt}, set conservatively within the 
manufacturer-specified range of \SIrange{2.75}{4.10}{\volt} \cite{ueki_development_2012} to protect cells whose 
degradation state is unknown.
As individual cells approach these thresholds, the maximum current of their module is progressively reduced (derating)
until the module drops out.
The system halts once the number of active modules falls below the minimum required to form the AC waveform.
Data were acquired from each module's onboard Cell Monitoring Unit (CMU) via the system's telemetry infrastructure.
Cell voltages were logged at \SI{10}{\second} intervals, module current and voltage at \SI{1}{\second} intervals,
and module temperature at \SI{15}{\second} intervals
(Figures~S1 and S2 in the supplementary material show the recorded data for all modules).
These nominal intervals are not always met (Figure~S3), since the data were recorded by the system's telemetry
during normal operation rather than by dedicated measurement equipment.
For module P1M9 (Phase~1, Module~9), an external current probe was used in addition to the onboard sensors,
enabling independent Coulomb counting and providing a reference for evaluating SOC estimation accuracy.

The dataset exhibits heterogeneity across cells and modules, as illustrated in Figure~\ref{fig:dataset}.
All modules show voltage deviations between cells (Figure~S1), most evident near the end of discharge,
where the OCV curve is steeper and individual cells reach voltage limits earlier.
Two cells, P1M6C1 and P3M5C12 (cell 1 of P1M6 and cell 12 of P3M5), diverge strongly from the rest of their modules, 
each alone constraining the usable capacity of its module (see also P3M5 in Figure~\ref{fig:dataset}).
The charge throughput over the window reflects this: most modules reach \SIrange{217}{238}{\Ah},
whereas P1M6 and P3M5 reach only \SI{95}{\Ah} and \SI{129}{\Ah}.
The raw data alone, however, cannot distinguish how much of the observed voltage spread originates
from SOH differences between cells, and how much from SOC imbalance.

\subsection*{Battery model and parametrization}

We combine a first-order ECM with recursive Gaussian process regression (RGP-ECM) to reconstruct the OCV curve and extract the ECM parameters from field operation data alone.
Each cell is characterized individually, using only its own measurements.
The battery state and all model parameters are estimated simultaneously via an extended Kalman filter (EKF), producing physically interpretable estimates together with their uncertainties.

Two sets of models are fitted: one at the cell level, using individual cell voltage measurements (324 cell models),
and one at the module level, using only the module terminal voltage (27 lumped-module models).
Both model sets share the same structure: an OCV source, an ohmic resistance $R_0$,
and an RC branch with resistance $R_1$ and time constant $\tau$ to capture transient behavior.
OCV and $R_1$ are each modeled as GPs, capturing their dependence on SOC.
Since cell capacity and initial SOC are unknown, the model is parametrized on cumulative charge $q$,
the integrated charge since the start of the measurement window, rather than normalized SOC.
The ohmic resistance $R_0$ and the time constant $\tau$ are estimated as scalars, 
neglecting their comparatively weak SOC dependence.
Temperature varies across modules (Figure~S2),
introducing differences in measured resistance unrelated to cell degradation.
All resistance parameters are therefore corrected via an Arrhenius factor estimated jointly with the other parameters.

\begin{figure}
    \centering
    \includegraphics[width=\textwidth]{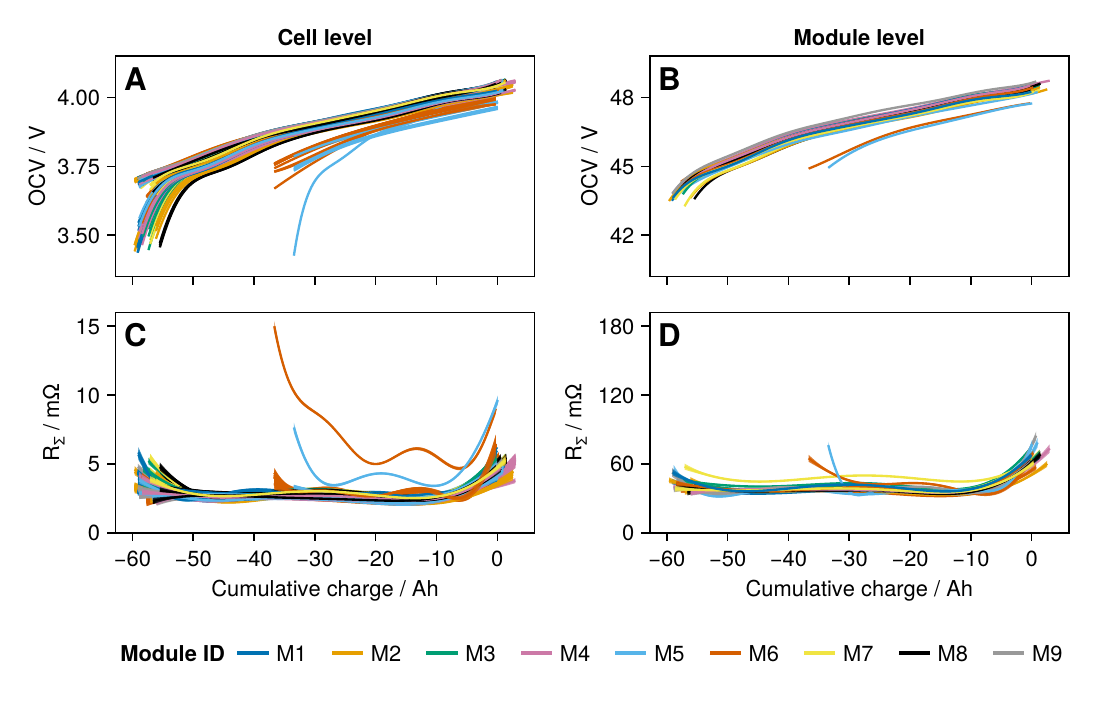}
    \caption{Reconstructed OCV curves and steady-state resistance $R_\Sigma$ at both model levels.
    (A, C) All 324 cell models. (B, D) All 27 module models.
    Curves are drawn against cumulative charge, colored by module position M1--M9 within each phase.
    Cumulative charge is counted from the start of the measurement window, so negative values indicate net discharge.
    $R_\Sigma$ is evaluated at the reference temperature of \SI{25}{\degreeCelsius}
    using the fitted Arrhenius factor.
    All cells show the same characteristic OCV shape, but each covers a different segment,
    depending on its capacity and initial SOC.
    Each module curve is fitted to twelve cells at once, so the variation within the module is lost.}
    \label{fig:ecms}
\end{figure}

\begin{figure}
    \centering
    \includegraphics[width=\textwidth]{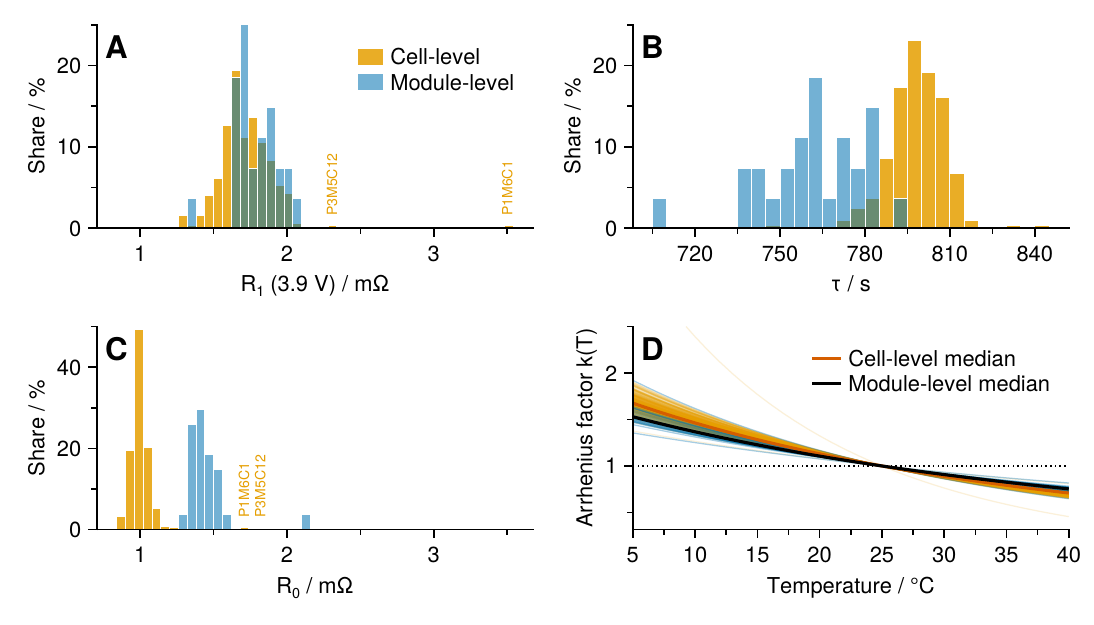}
    \caption{Fitted ECM parameters of all 324 cells (orange) and 27 modules (blue).
    Distributions of (A) the RC-branch resistance $R_1$ evaluated at an OCV of \SI{3.9}{\volt},
    (B) the time constant $\tau$, and (C) the ohmic resistance $R_0$.
    (D) Arrhenius correction factor $k(T)$, one line per cell and module,
    with the medians as thick lines and $k(T) = 1$ at the reference temperature of \SI{25}{\degreeCelsius}.
    Resistances are corrected to that temperature
    and module values divided by the twelve series-connected cells,
    so both levels are directly comparable.}
    \label{fig:ecm-parameters}
\end{figure}

Across all 324 cells, the reconstructed OCV curves share the same qualitative shape,
characteristic of an LMO/graphite cell (Figure~\ref{fig:ecms}A).
The curves differ, however, in the window each cell covers, in both position and span along the charge and voltage axes.
This variation indicates a spread in SOH and SOC imbalance between cells,
although the contributions of each cannot be separated from the reconstructed curves directly.
The spread is most evident in the modules containing the cells with distinctive behavior (P1M6C1, P3M5C12).
The OCV curve of P3M5C12 spans most of the operational voltage range.
The cell therefore reaches both voltage limits before any other cell in the module,
confining the remaining cells to only a fraction of their own voltage window.
The two cells stand out in the reconstructed resistances as well.
Figure~\ref{fig:ecms}C shows the cell-level steady-state resistance $R_\Sigma$, 
corrected to a reference temperature of \SI{25}{\degreeCelsius} via the Arrhenius fit:
\begin{equation}
    R_\Sigma(q) = R_0(T=\SI{25}{\degreeCelsius}) + R_1(q, T=\SI{25}{\degreeCelsius})
\end{equation}
Most cells cluster within a moderate spread and show only a mild charge dependence:
$R_\Sigma$ increases toward both ends of each cell's operating window, 
consistent with the expected behavior at low and high SOC 
\cite{waag_experimental_2013,tran_comprehensive_2021}.
The two outlier cells display substantially higher resistance, with a more pronounced charge dependence.

At the module level, the terminal voltage alone cannot resolve the cell-level heterogeneity 
(Figure~\ref{fig:ecms}B and D).
Expressed per cell equivalent, the module OCV curves span a narrower voltage range than the individual cell curves, since operation stops once any cell in the module reaches its voltage limit.
Modules containing the outlier cells remain identifiable only from their reduced charge coverage.
The module-level $R_\Sigma$ shows the same effect, with the variation across the twelve cells absorbed into a single 
aggregate curve (Figure~\ref{fig:ecms}D).

Figure~\ref{fig:ecm-parameters} shows the distributions of the fitted ECM parameters $R_1$, $\tau$, $R_0$,
and the Arrhenius factor over all 324 cells and 27 modules,
with the module resistances expressed per cell equivalent for direct comparison.
Since $R_1$ varies with charge, it is evaluated where the OCV reaches \SI{3.9}{\volt},
which serves as a common reference in the absence of a known SOC.
Both levels agree over the bulk of the $R_1$ distribution, but the lumped-module models do not reproduce the extreme 
values of P1M6C1 and P3M5C12, which reach 2.1 and 1.4 times the population median (Figure~\ref{fig:ecm-parameters}A).
The same holds for $R_0$, where both outlier cells reach about 1.7 times the population median
(Figure~\ref{fig:ecm-parameters}C).
Unlike $R_1$, however, the two distributions are also offset, with the module-level median 
approximately \SI{0.4}{\milli\ohm} above the cell-level one.
The offset likely reflects the resistance of the module's contacts and busbars,
which the cell voltage measurement does not include.

The fitted Arrhenius factors are comparable at cell and module level.
Over \SIrange{15}{35}{\degreeCelsius}, which spans the operating temperatures in the dataset,
the fitted dependence changes the resistance by about \SIrange{50}{60}{\percent},
more than the spread of the resistances across the population (Figure~\ref{fig:ecm-parameters}D).
Without this correction, differences in operating temperature would be indistinguishable
from differences in degradation.

Each module-level resistance averages over the twelve cells in series,
so a single degraded cell shifts it only slightly.
The module-level distributions therefore describe the average condition of each module
rather than that of its weakest cell.

\subsection*{SOH heterogeneity and SOC imbalance}

The reconstructed OCV curves are parametrized on cumulative charge $q$, each starting from an unknown initial SOC.
To disentangle SOH differences from SOC imbalance, they must be aligned to a common SOC axis.
Since each cell covers a partial window of the full OCV range, however, 
neither capacity nor initial SOC can be read from any single curve.
We instead scale and shift all curves simultaneously to minimize disagreement over shared voltage ranges,
recovering capacity and initial SOC for each cell from the joint fit. 

\begin{figure}
    \centering
    \includegraphics[width=\textwidth]{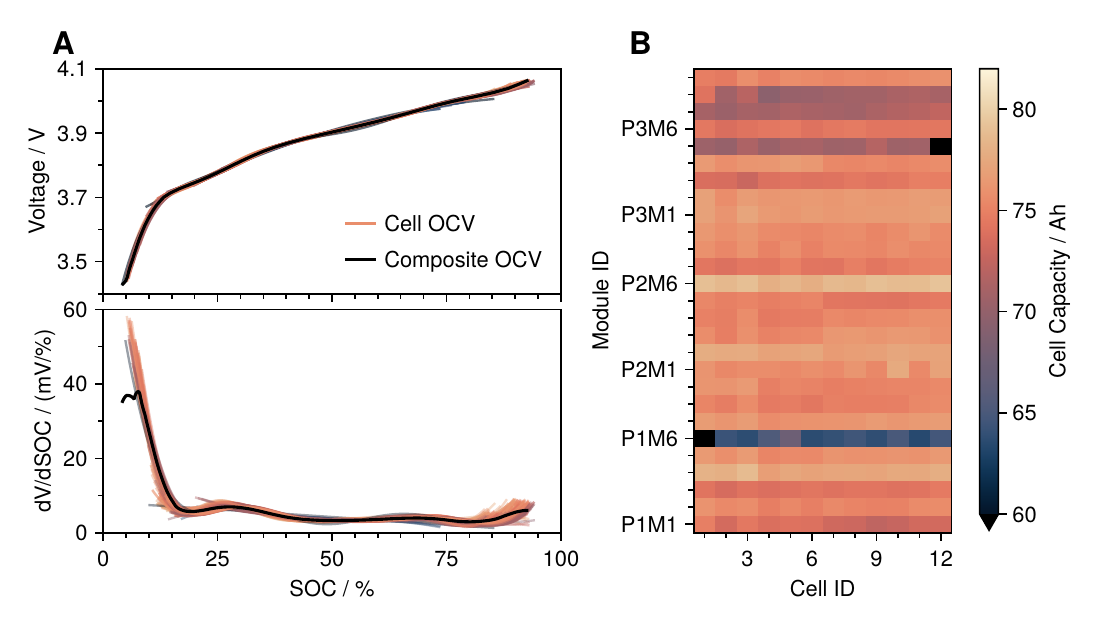}
    \caption{Alignment of the reconstructed OCV curves and the resulting cell capacities.
    (A) The 324 reconstructed OCV curves aligned on a common SOC axis,
    together with the composite OCV curve averaged from them (black),
    and the corresponding \dvdsoc~below.
    (B) Estimated capacity of every cell, arranged by cell position within each module.
    Curves in (A) and tiles in (B) are colored by the estimated cell capacity on a shared scale,
    with capacities below \SI{60}{\Ah} drawn in black in (B).
    Capacity varies less within a module than between modules,
    with P1M6C1 and P3M5C12, the black tiles in (B), as the exceptions.}
    \label{fig:cell-soh}
\end{figure}

Figure~\ref{fig:cell-soh}A shows the resulting alignment of the 324 cell OCV curves.
Also shown is the \textit{composite} OCV curve, assembled by averaging the aligned partial curves into
a single reference.
Spanning \SIrange{3.43}{4.06}{\volt}, the composite covers a wider voltage range than any individual cell,
whose median voltage span over the measurement window is only \SI{0.36}{\volt}.
Despite being reconstructed independently, the aligned OCV curves agree with the composite
to a median per-cell RMSE of \SI{1.95}{\milli\volt}, with a 95th percentile of \SI{3.97}{\milli\volt}.
P1M6C1 and P3M5C12 show the highest RMSE in the population,
at \SI{9.0}{\milli\volt} and \SI{7.0}{\milli\volt} respectively.
The \dvdsoc{} of the composite OCV shows two distinct peaks and an intermediate plateau,
which most aligned cell curves reproduce.

The estimated cell SOH shows substantial cell-to-cell variation (Figure~S4).
Most cells are tightly grouped around a median of \SI{75.4}{\percent}, with \SI{50}{\percent} of cells
between \SIrange{74.3}{76.2}{\percent}, but the distribution has a pronounced lower tail.
Figure~\ref{fig:cell-soh}B reveals the structure behind this variation:
cells within the same module tend to share similar SOH levels,
with within-module standard deviations typically below \SI{1}{\percent},
while the module average differs noticeably between modules.
In P1M6 and P3M5, however, a single cell departs sharply from the rest of its module:
P1M6C1 at \SI{47.8 \pm 1.4}{\percent} and P3M5C12 at \SI{38.0 \pm 0.4}{\percent},
\SI{16}{\percent} and \SI{32}{\percent} below the second-lowest cell in their respective modules
(uncertainties denote one standard deviation).
The remaining cells of P1M6 also fall below the population, but their estimates are less certain:
P1M6C1 confines the module to a narrow voltage window, which leaves the other cells' curves
less constrained in the alignment.

\begin{figure}
    \centering
    \includegraphics[width=\textwidth]{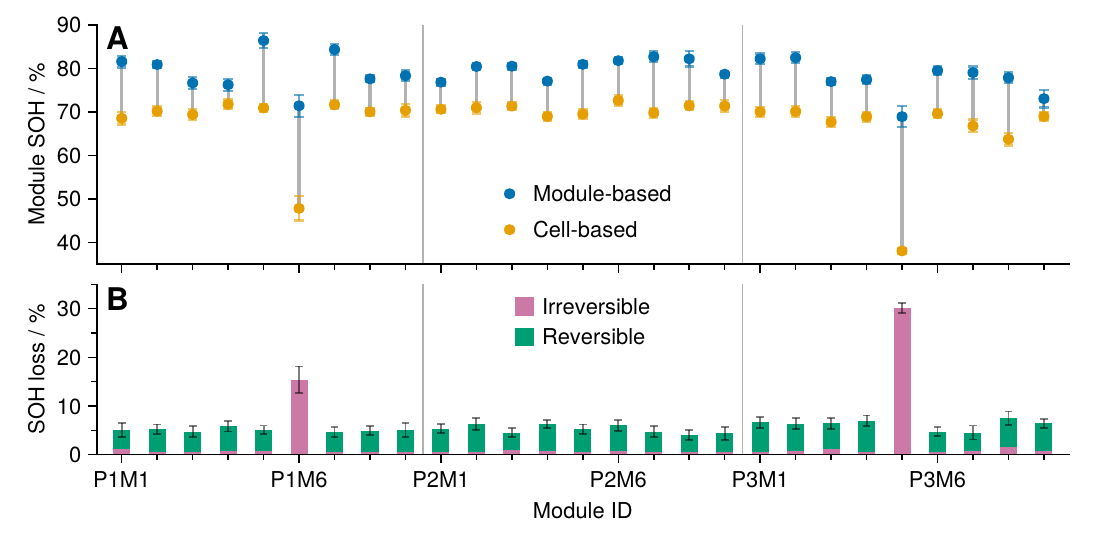}
    \caption{Estimated module SOH from the cell models and the lumped-module model.
    (A) SOH of each module aggregated from its twelve cell estimates (orange)
    and estimated directly by the lumped-module model (blue),
    with the gray connector marking the gap between them.
    (B) The corresponding capacity loss per module, split into a reversible part,
    caused by SOC imbalance and recoverable by balancing,
    and an irreversible part set by the spread in cell capacity.
    Whiskers give the $\pm 2\sigma$ uncertainty of each estimate.
    The lumped-module model reports a consistently higher SOH than the cell-based aggregate,
    most severely in P1M6 and P3M5.}
    \label{fig:module-soh}
\end{figure}

In addition to cell capacity, the OCV curve alignment recovers each cell's initial SOC, 
revealing a systematic SOC imbalance of \SIrange{6}{11}{\percent} between the lowest and highest cell of each module.
The imbalance follows the same pattern in every module, with cells C1, C2, and C3 consistently
below the remaining cells, which points to a systematic origin rather than random drift.
These cells, however, show no corresponding deviation in capacity.
The resulting SOC offset directly reduces each module's available capacity (Figure~\ref{fig:module-soh}B), 
leaving \SIrange{3.5}{6.5}{\Ah} inaccessible in most modules.
Although most of this inaccessible capacity could in principle be recovered through cell balancing,
balancing is triggered only by extended rest periods, which the measurement window does not contain.
Even after balancing, the SOH heterogeneity between cells leaves a residual \SIrange{0.5}{1.6}{\Ah} per module 
inaccessible, which cannot be recovered.
In P1M6 and P3M5, however, the outlier cell in each module is so severely degraded
that it constrains both the charge and discharge limits.
The inaccessible capacity of \SI{15.4}{\Ah} and \SI{30.1}{\Ah} is therefore entirely irreversible,
and cell balancing provides no benefit.
By capturing this cell-level heterogeneity, we evaluate the usable capacity of each module,
set by the first cell to reach either voltage limit (Figure~\ref{fig:module-soh}A).
The module SOH is determined by the smallest chargeable and dischargeable capacity among its cells
(subsection OCV curve alignment and SOH estimation in Methods).
For 25 of the 27 modules, the resulting SOH ranges from \SIrange{64}{73}{\percent},
consistently below the mean cell SOH and dominated by the recoverable SOC imbalance.
In P1M6 and P3M5, the module SOH collapses to that of the single outlier cell, 
far below the range of the remaining modules.

Without access to individual cell voltages, the lumped-module model cannot resolve the intra-module heterogeneity
that determines usable capacity.
Applying the same OCV curve alignment to the 27 lumped-module models yields a direct SOH estimate for each module.
The lumped-module and cell models achieve comparable fit quality, with a median RMSE of \SI{2.28}{\milli\volt}
and \SI{1.95}{\milli\volt} per cell equivalent respectively.
The fit residuals of P1M6 and P3M5, at \SI{2.76}{\milli\volt} and \SI{2.80}{\milli\volt} per cell equivalent,
lie only slightly above this median.
The fit quality alone does not, however, guarantee accurate SOH estimation.
The lumped-module SOH estimates exceed the cell-aggregated values by \SIrange{4}{16}{\percent} for 25 of the 27 modules 
and by \SI{24}{\percent} and \SI{31}{\percent} for P1M6 and P3M5 (Figure~\ref{fig:module-soh}A).
The larger divergence in P1M6 and P3M5 arises because the outlier cell, despite setting the usable capacity
of the entire module, contributes only a fraction of the module terminal voltage signal.
How much the lumped-module model overestimates usable capacity depends directly on
the degree of intra-module heterogeneity.

\subsection*{Validation against reference measurements}

We validate the reconstructed OCV curves, together with the estimated cell capacity and the SOC imbalance,
against a dedicated reference measurement.
For the reference measurement, a dedicated data logger recorded the CMU measurements
of the nine modules of phase P3 locally.
Module P3M5 is excluded from the comparison since its faulty cell (P3M5C12)
was fixed between the two experiments (see Discussion), so the module state differs between the two measurements.
The validation therefore covers eight of the 27 modules with 96 cells, roughly a third of the system.
Apart from the two divergent cells,
the validated subset spans the capacity range and the SOC imbalance observed across the system
and is therefore representative of the population.

The reference measurement consists of one discharge and one subsequent charge,
performed by the system itself at a constant power setpoint of \SI{3.8}{\kilo\watt},
corresponding to cell currents of roughly C/25.
For this measurement the lower cell voltage limit was reduced to \SI{3.2}{\volt},
extending the coverage of the OCV curve into its steep low-SOC region.
Cell voltage and module current come from the same onboard CMU sensors as in the cycling experiment,
and charge is obtained by integrating the module current.
The MMC modulation alternates each module between load pulses and rest intervals at zero current.
From each rest interval we keep the final sample, where the voltage has relaxed closest to the OCV,
and assemble these samples into one charge and one discharge curve per cell.
The rest intervals last about \SI{60}{\second}, short compared to the slow relaxation of the cells,
so both curves retain residual polarization in addition to OCV hysteresis.
We therefore take the pseudo-OCV of each cell as the average of its charge and discharge curve,
which cancels the symmetric part of both effects.

\begin{figure}
    \centering
    \includegraphics[width=0.95\textwidth]{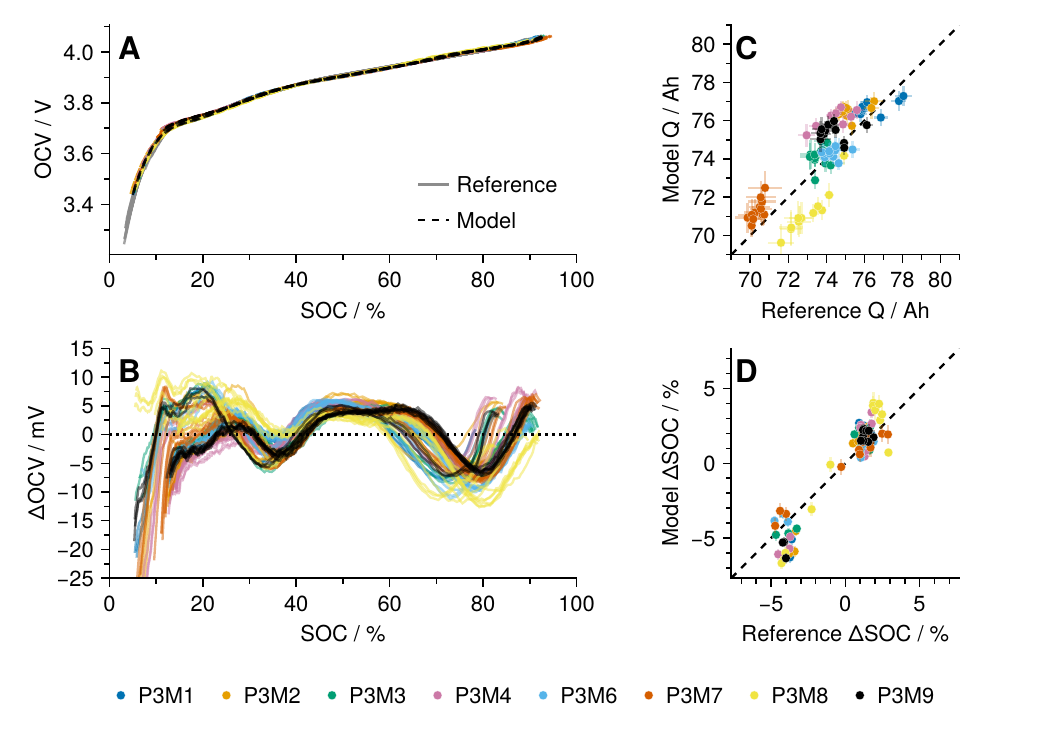}
    \caption{Validation of the reconstructed OCV curves, estimated cell capacities, and SOC imbalance.
    The model estimates are compared with the reference measurement
    for the 96 cells of the eight validated modules.
    (A) Reconstructed OCV curves (dashed, colored by module) and reference pseudo-OCV curves (gray),
    aligned pairwise and normalized to SOC.
    (B) Voltage residual between the aligned curves of each cell.
    (C) Estimated cell capacity, model against reference.
    (D) Estimated SOC imbalance, the deviation of each cell's initial SOC from its module mean,
    model against reference.
    Error bars give $\pm 1\sigma$ uncertainty intervals.}
    \label{fig:ocv-validation}
\end{figure}

To compare the reconstructed OCV curves with the pseudo-OCV references,
we align each pair by a constant shift in charge, since the charge origin of the two experiments is arbitrary.
Both the shape and the capacity scale of the curves are preserved by this alignment (Figure~\ref{fig:ocv-validation}A).
We quantify the agreement by the RMSE of the voltage residual between the aligned curves,
evaluated over each cell's own overlap window.
The median RMSE across the 96 cells is \SI{4.0}{\milli\volt},
with a 90th percentile of \SI{8.5}{\milli\volt} and a maximum of \SI{14.3}{\milli\volt} (Figure~S5).
The largest deviations occur in the cells with the lowest SOC, whose curves cover the steep knee of the OCV curve,
where small charge errors translate into large voltage residuals.
The residual shares a common shape across all cells, following a wave-like pattern with an amplitude of 
up to \SI{10}{\milli\volt} over the mid-SOC range (Figure~\ref{fig:ocv-validation}B).
This systematic residual may originate from a combination of the model reconstruction and the reference itself,
where the pseudo-OCV can be skewed by the asymmetric part of polarization and hysteresis.

From the pseudo-OCV curves we also extract a reference capacity and initial SOC for each cell,
using the same curve alignment as for the reconstructed curves.
The same SOC-voltage anchor points as for the model estimates scale the reference capacities to the full voltage window.
The SOC imbalance is compared as the deviation of each cell from its module mean,
since the initial SOC differed between the two experiments.

Figure~\ref{fig:ocv-validation}C compares the estimated cell capacities with their reference values.
In absolute terms, the two agree closely, with an RMSE of \SI{1.25}{\Ah}.
However, the capacity estimates of the cells within a module tend to deviate from the reference in a common direction,
most visibly in P3M8.
A plausible origin for this bias is the current measurement, which all twelve cells of a module share.
An error in the integrated charge then shifts their capacities together, in either of the two experiments.
When each cell's capacity is instead expressed as the deviation from its module mean,
the estimates agree to an RMSE of \SI{0.55}{\Ah}.
This remaining error lies within the uncertainties of both the model and the reference estimates.

Figure~\ref{fig:ocv-validation}D compares the SOC imbalance between the two datasets.
Model and reference capture the same pattern:
cells C1--C3 typically lie \SIrange{4}{5}{\percent} below the module mean,
while the remaining cells sit slightly above it.
The cell deviations agree with an RMSE of \SI{1.1}{\percent} and a maximum difference of \SI{2.6}{\percent}.
These differences are larger than the uncertainties of both estimates.
However, the experiments took place about seven weeks apart,
during which the relative charge of the cells could have drifted,
for example through uneven self-discharge.

\subsection*{Voltage prediction accuracy and SOC estimation}

\begin{figure}
    \centering
    \includegraphics[width=\textwidth]{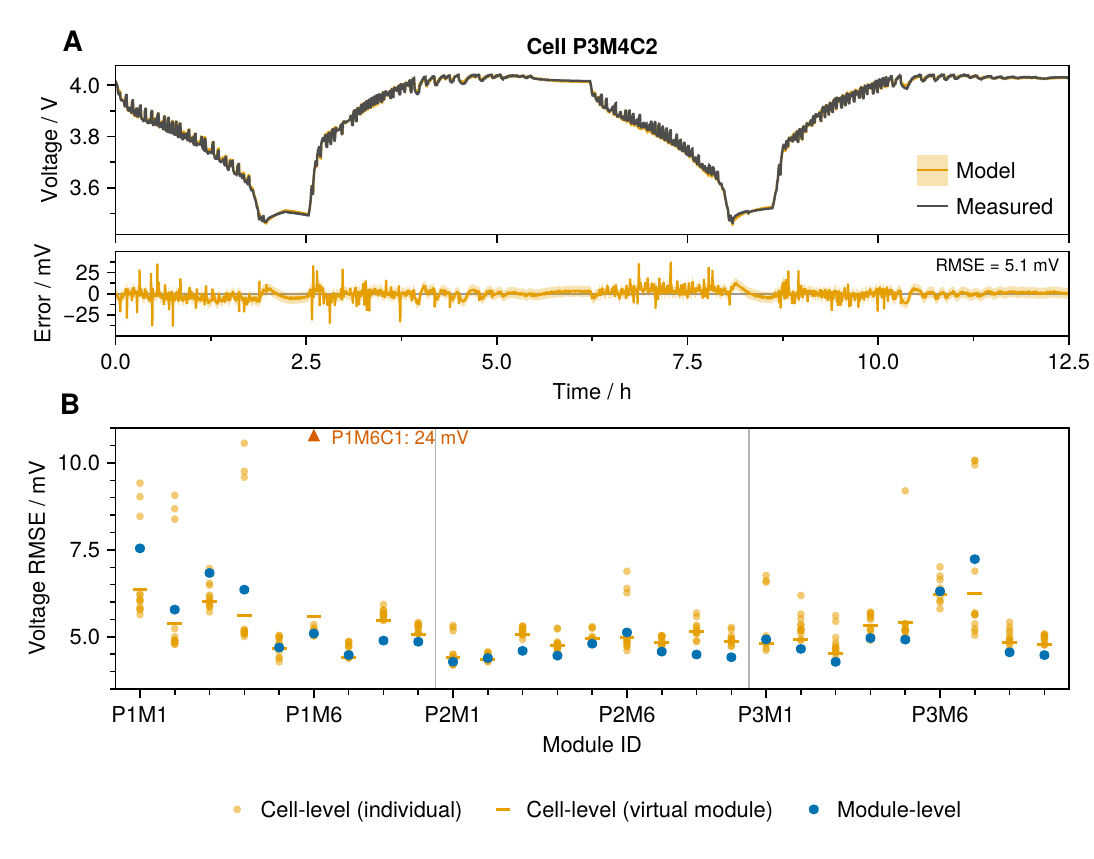}
    \caption{Voltage prediction accuracy of the fitted models, run in open loop without correction steps.
    (A) Measured (gray) and predicted (orange) terminal voltage of cell P3M4C2
    over the measurement window, with the prediction error below.
    (B) Voltage RMSE of every cell (orange points), of the cell models combined per module (orange ticks),
    and of the lumped-module models (blue), all per cell equivalent.
    P1M6C1 lies above the axis range and is labeled in place.
    Both model sets reproduce the module voltage with comparable accuracy.}
    \label{fig:model-voltage-accuracy}
\end{figure}

The parametrized models can also serve as online state estimators: fixing the parameters converts the joint estimator into a pure state observer that tracks the charge state by correcting its voltage prediction against measurements.
The accuracy of the state estimates depends on how well the model predicts the voltage response. 
We characterize this by running both the cell and lumped-module models in open loop, without measurement updates.
Figure~\ref{fig:model-voltage-accuracy}A shows cell P3M4C2 as a representative example.
The model closely tracks the measured voltage over the full \SI{12.5}{\hour} window.
Across the full population, the median per-cell RMSE is \SI{5.09}{\milli\volt}, 
with \SI{95}{\percent} of cells below \SI{6.89}{\milli\volt}, P1M6C1 being the sole outlier at \SI{24.0}{\milli\volt}.
When cell-model predictions are combined per module, the resulting virtual voltage has 
a median RMSE of \SI{4.98}{\milli\volt} per cell equivalent, against \SI{4.80}{\milli\volt}
for the lumped-module model (Figure~\ref{fig:model-voltage-accuracy}B).
The lumped-module model thus matches the module voltage at least as well as the cell models,
yet it substantially overestimates SOH.
Individual cell differences contribute only a small fraction of the aggregate terminal voltage signal, 
so the lumped-module model fits it accurately without resolving the internal distribution, 
and therefore cannot determine which cell first approaches its voltage limit.

With measurement updates restored, the EKF tracks the charge of each cell in real time.
We validate its charge estimate against the independent Coulomb-counting reference 
from the external current probe on P1M9.
For the twelve cells of the module, the estimation error stays below \SI{1}{\Ah} for every cell over the full \SI{12.5}{\hour} window (Figure~S6).
To test the case of an unknown initial state and a drifting sensor,
we additionally ran a synthetic scenario with a charge offset and a current bias injected into the onboard current.
Coulomb counting with that current then drifts substantially,
whereas the EKF corrects both faults through the voltage feedback (Figure~S7).
These cell-level trajectories are the basis for the module SOC estimates that follow.

\begin{figure}[tbp]
    \centering
    \includegraphics[width=\textwidth]{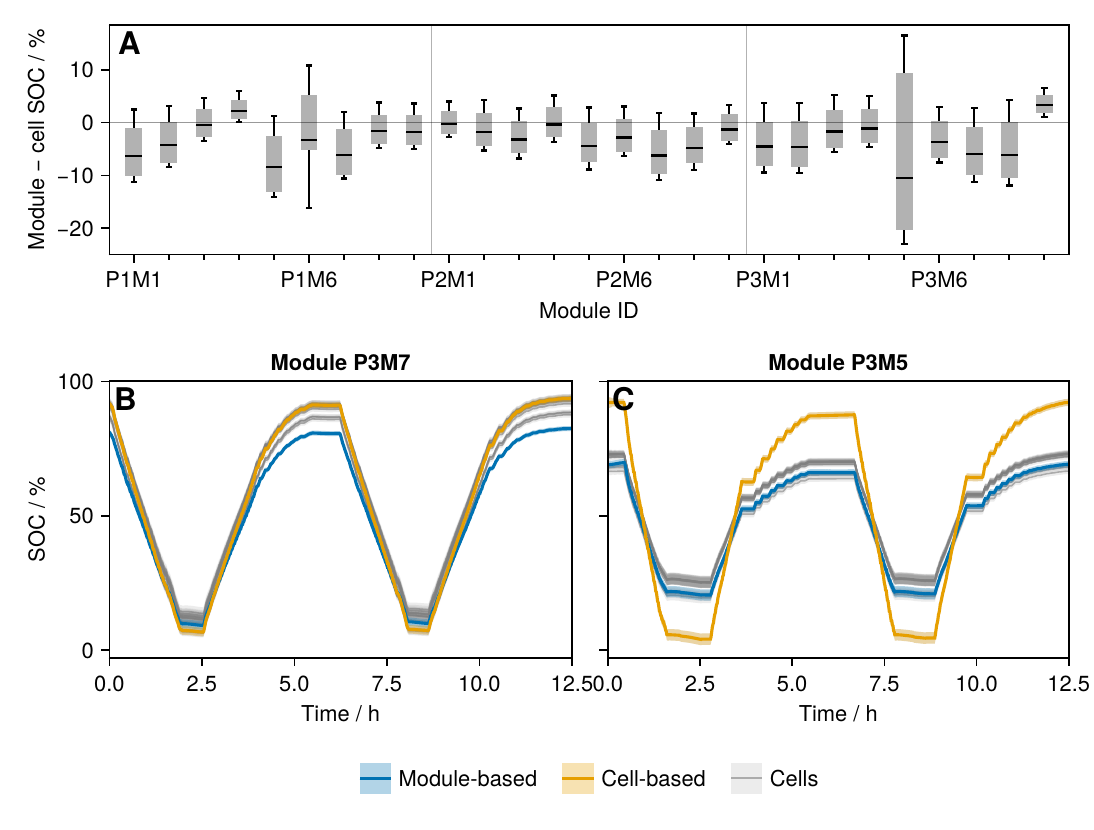}
    \caption{SOC estimation from the cell models and the lumped-module model.
    (A) Difference between the lumped-module and cell-based SOC estimates,
    each box summarizing the distribution over all time steps of one module.
    (B, C) SOC trajectories of two example modules, P3M7 and P3M5,
    showing the twelve individual cell estimates (gray), their aggregate (orange),
    and the lumped-module estimate (blue), all with $\pm 2\sigma$ uncertainty bands.
    The lumped-module estimate follows the module's average behavior rather than the limiting cell,
    and therefore overstates the usable range.}
    \label{fig:soc-error}
\end{figure}

To reduce all twelve cells in a module to a single SOC value, we aggregate their charge trajectories
(subsection Simulation and SOC estimation in Methods).
The aggregation preserves the limiting cell in each direction, which an average over the twelve would obscure.
In P3M7, as in most modules, this estimate closely tracks the discharge-limiting cell at low SOC
and the charge-limiting cell at high SOC, blending between the limiting cells through the mid-range
(Figure~\ref{fig:soc-error}B).
Which cell limits charge or discharge is not fixed over the course of the run:
in a module with evenly degraded cells such as P3M7, this role shifts between several cells.
In the two outlier modules, by contrast, one cell sets both limits, so the module SOC estimate
reduces to the SOC of that single cell.
Figure~\ref{fig:soc-error}C shows module P3M5 as an example.

We compare this cell-based estimate against the direct estimate of the lumped-module model,
obtained by applying the same EKF to the lumped-module ECM instead of the twelve cell models.
In P3M7, the lumped-module estimate reaches \SI{82.5 \pm 0.5}{\percent} SOC at the charge peak,
against \SI{93.7 \pm 0.6}{\percent} for the cell-based estimate, underestimating SOC by \SI{11.2 \pm 0.7}{\percent}.
At the discharge minimum, it instead overestimates SOC by \SI{2.5 \pm 1.1}{\percent}.
Across all 27 modules, the maximum deviation ranges from \SI{4.0 \pm 0.7}{\percent} (P2M1) 
to \SI{23.1 \pm 1.2}{\percent} (P3M5), as shown in Figure~\ref{fig:soc-error}A.
As with the discrepancies in SOH estimation, this pattern of overestimating SOC at low charge
and underestimating it at high charge follows from the lumped-module model tracking the average of the cells
rather than whichever cell is currently closest to its limit.

\section*{DISCUSSION}
Second-life battery systems are difficult to characterize,
due to cell-to-cell variation in capacity and resistance caused by years of uneven degradation.
Furthermore, model-based SOC and SOH estimation requires a parametrized cell model, and in particular an OCV curve,
which is not available for cells of unknown provenance and degradation state.
We adopt a battery model that embeds Gaussian-process components in an ECM.
An extended Kalman filter recursively estimates its state and parameters,
reconstructing the OCV curve from the voltage, current, and temperature measured during field operation.
Each cell covers only part of its OCV curve during the measurement window.
Aligning the reconstructed curves across the population recovers the capacity and initial SOC of every cell.
In the case study, the model is fitted individually to each of the 324 cells from a \SI{12.5}{\hour} measurement window.
In open loop, the fitted models predict the measured voltage to a median RMSE of \SI{5.09}{\milli\volt}.
Against a dedicated reference measurement, the reconstructed OCV curves agree to a median RMSE of \SI{4.0}{\milli\volt},
the estimated cell capacities to \SI{1.25}{\Ah}, and the SOC imbalance to \SI{1.1}{\percent}.

Across the 324 cells, the estimated capacities vary substantially, 
with two cells falling far below the rest of the population.
The alignment also reveals a systematic SOC imbalance in every module.
Aggregating the cell estimates yields the usable capacity and effective SOC of each module,
determined by whichever cell reaches its voltage limit first.
For comparison, we apply the same procedure to a lumped-module model, 
fitting a single ECM to each module and estimating its SOH and SOC without access to the individual cell voltages.
Both the cell models and the lumped-module ECM predict voltage with comparable accuracy.
However, since the lumped-module model follows the average condition of its cells, 
it cannot identify the limiting cells,
and consequently overestimates SOH by up to \SI{31}{\percent} and SOC by up to \SI{23}{\percent}.
Voltage accuracy alone is therefore no indication that the internal state of a module has been resolved.

From the raw cell voltage data alone, P1M6C1 and P3M5C12 stand out, deviating clearly from the rest of their modules.
Our characterization shows that both have roughly half the capacity and twice the resistance of the population median. 
Since each logical cell consists of two physical cells in parallel, the loss of one of them would produce this behavior.
Inspection after the study traced both faults to the busbar contacts, which were subsequently fixed.
Because the model is parametrized for every cell from field data alone, 
it quantifies how much usable capacity remains when such a fault is present, 
and how the fault constrains system operation.

\subsection*{Limitations of the study}

The OCV curve alignment recovers the capacity of each cell over the voltage range the data covers.
In practice, battery systems are generally operated within conservative limits,
so field data rarely reaches the rated voltage window.
Obtaining SOH relative to the nominal capacity therefore requires an extrapolation beyond the measured range,
which is prone to systematic error.
The validation is subject to similar constraints:
the reference measurement stems from the same onboard sensors and shares the extrapolated reference points,
leaving the absolute capacity scale unverified.
A conclusive assessment would require characterizing cells individually in the laboratory.

In this study, the parametrization rests on a short measurement window 
and returns a snapshot of the system state and parameters.
A natural next step is long-term monitoring,
tracking resistance increase and capacity loss over months and years \cite{aitio_learning_2025}.
Since the reconstructed curves already resolve the characteristic \dvdsoc{} features of the LMO/graphite cell,
tracking them over time would additionally give access to degradation mode analysis
\cite{zhou_learning_2025, figgener_degradation_2024}.

The cell-level models require every cell voltage to be transmitted, stored, and processed.
Over years of operation, the cost of telemetry and computing infrastructure can become substantial.
Cheaper approaches, such as tracking only the limiting cells,
offer a trade-off between cost and accuracy that is worth exploring \cite{li_interval_2017}.

\section*{METHODS}
\subsection*{Hybrid battery model}

We reconstruct the OCV curve and identify the remaining ECM parameters from field data,
since no pre-measured curve is available for second-life cells.
The OCV curve and an SOC-dependent resistance are modeled as GPs,
which recover their shape directly from the data instead of fitting them to a predetermined equation.
To estimate the GPs together with the ECM parameters as measurements arrive, 
we follow the recursive formulation of GP regression introduced by Huber \cite{huber_recursive_2014}.
The recursive GP (RGP) represents each function by a fixed set of $m$ basis points,
a vector of the corresponding function values, and their covariance matrix, which a Kalman filter updates.
The function can then be evaluated at any input from these basis point values,
with an uncertainty estimate that reflects how well the observed data support the prediction.
In contrast with standard GP regression, which retains the full history of observations,
the RGP carries the fitted function in its state,
allowing it to update online and to handle large datasets.
Because each function is reduced to a finite vector of values,
the curves and the ECM parameters can be estimated in a single filter.
Note S1 in the supplementary material gives a detailed mathematical description of GP regression and its recursive formulation.

We model each cell (and battery module) with a first-order ECM:
an OCV source, an ohmic resistance $R_0$, and an RC branch with resistance $R_1$ and time constant $\tau$.
The OCV curve and $R_1$ are modeled as \textit{functional parameters}, each represented by a recursive GP.
As a simplifying assumption, we treat $R_0$ and $\tau$ as constant with respect to the SOC.
While both could be extended and modeled as GPs to also capture SOC dependency,
too many degrees of freedom in the model may cause an ill-conditioned estimation problem 
without necessarily improving the model accuracy \cite{hu_comparative_2012,grandjean_structural_2017}.
$\tau$ is generally the ECM parameter to which terminal voltage is least sensitive
\cite{zhao_global_2016,lai_parameter_2020},
and some studies report a comparatively small SOC dependence for $R_0$ relative to $R_1$ 
\cite{andre_characterization_2011,waag_experimental_2013,tran_comprehensive_2021}.
Rather than normalized SOC, both the OCV curve and $R_1$ are parametrized on cumulative charge 
$q(t) = \int_0^{t} i(t') \, \mathrm{d}t'$, 
since neither the cell's capacity nor its initial SOC is known ahead of the fit.
The mapping back to SOC is recovered only once each cell's capacity and initial SOC are identified from the ensemble of reconstructed OCV curves, described in the subsection OCV curve alignment and SOH estimation.

Both $R_0$ and $R_1$ are also dependent on temperature,
which varies considerably both across modules and over the course of the measurement window.
We assume both $R_0(T) = R_0 \, k(T)$ and $R_1(q,T) = R_1(q) \, k(T)$ share the same correction factor
$k(T)$ with Arrhenius temperature dependence,
\begin{equation}
    k(T) = \exp\left(\kappa\left(\frac{1}{T} - \frac{1}{T_0}\right)\right)
\end{equation}
with reference temperature $T_0 = \SI{25}{\degreeCelsius}$, 
and where $\kappa$, an activation-energy-like coefficient, is estimated jointly with the other parameters.
Although $R_0$ and $R_1$ likely have different activation energies in practice \cite{ludwig_determination_2022}, both respond to the same
temperature change in the same voltage signal, so separate values of $\kappa$ are difficult to resolve \cite{barat_system_2026}.
We therefore share $\kappa$ between them.

\begin{figure}
    \centering
    \includegraphics[width=\textwidth]{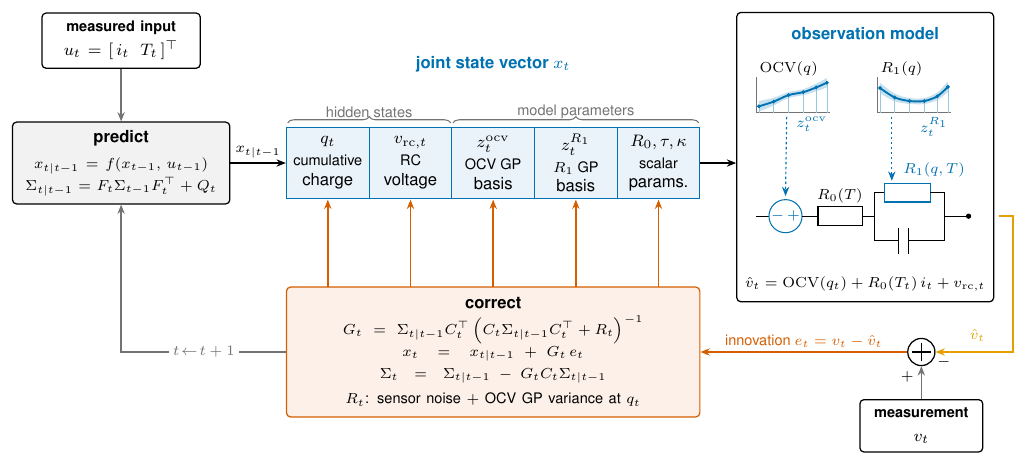}
    \caption{
        Joint state and parameter estimation with the RGP-ECM.
        The state vector $x_t$ carries the hidden states alongside the ECM parameters,
        so the EKF updates them jointly at each step.
        The GPs recover $\mathrm{OCV}(q)$ and $R_1(q)$ as functions of charge, while
        both resistances additionally depend on temperature through a shared
        Arrhenius factor.
        $F_t$ and $C_t$ are the Jacobians of the process model $f$ and the observation
        model $g$, evaluated at $x_{t-1}$ and $x_{t|t-1}$.
    }
    \label{fig:method-ekf}
\end{figure}

An EKF jointly estimates all the ECM parameters
and the hidden states $q$ and $v_\text{rc}$, combined into a single state vector (Figure~\ref{fig:method-ekf}),
\begin{equation}
    x_t = \begin{bmatrix} 
        q_t 
        & v_{\text{rc},t} 
        & z_t^\text{ocv} 
        & z_t^{R_1} 
        & R_{0,t} 
        & \tau_t 
        & \kappa_t 
    \end{bmatrix}^\top
\end{equation}
The basis-point values $z_t^\text{ocv}$ and $z_t^{R_1}$ of the recursive GPs $\mathrm{OCV}(q)$ and $R_1(q)$
are ordinary entries of this vector, alongside the scalar parameters $R_0$, $\tau$, and $\kappa$.

The state evolves from step to step with the model dynamics $x_t = f(x_{t-1}, u_{t-1}) + \eta_t$,
with process noise $\eta_t \sim \mathcal{N}(0, Q_t)$,
where $u_t = \begin{bmatrix} i_t & T_t \end{bmatrix}^\top$ is the input vector of measured current and temperature.
$f$ combines a distinct update rule for each state component: the internal charge accumulation
is updated with Coulomb counting,
\begin{equation}
    q_t = q_{t-1} + i_{t-1}\,T_s
\end{equation}
with $T_s$ the filter's time step ($T_s = \SI{1}{\second}$ in this case).
The RC-branch voltage evolves with the standard zero-order-hold discretization,
\begin{equation}
    v_{\text{rc},t} = v_{\text{rc},t-1} \, \exp(-T_s/\tau)
    + R_1(q_{t-1}, T_{t-1}) \, i_{t-1} \, \left(1-\exp(-T_s/\tau)\right)
\end{equation}
Every other state, both the GPs' basis-point values and the three scalars,
follows identity dynamics with zero process noise and is refined only through the correction step.

The correction step uses this prediction together with the observation model $y_t = g(x_t, u_t) + \epsilon_t$,
where $\epsilon_t \sim \mathcal{N}(0, R_t)$ is the sensor noise and $g(x_t,u_t)$ predicts the terminal voltage $\hat v_t$,
combining the OCV voltage source with the ohmic and dynamic voltage drops,
\begin{equation}
    \hat v_t = \mathrm{OCV}(q_t) + R_0(T_t)\,i_t + v_{\text{rc},t}
\end{equation}
$f$ and $g$ are nonlinear in $x_t$: $q$ enters through the recursive GPs $R_1(q)$ and $\mathrm{OCV}(q)$,
while $\tau$ and $\kappa$ enter through the exponential terms.
The filter therefore linearizes both functions at each step via their Jacobians,
computed via forward-mode automatic differentiation \cite{revels_forward_2016}:
\begin{equation}
    F_t = \left.\frac{\partial f}{\partial x}\right|_{x_{t-1}}, \qquad
    C_t = \left.\frac{\partial g}{\partial x}\right|_{x_{t|t-1}}
\end{equation}
The prediction step propagates the state and covariance,
\begin{equation}
    x_{t|t-1} = f(x_{t-1}, u_{t-1}), \qquad \Sigma_{t|t-1} = F_t \Sigma_{t-1} F_t^\top + Q_t
\end{equation}
and the correction step updates them using the Kalman gain,
\begin{align}
    G_t &= \Sigma_{t|t-1} C_t^\top \left(C_t \Sigma_{t|t-1} C_t^\top + R_t\right)^{-1} \\
    x_t &= x_{t|t-1} + G_t\left(y_t - g(x_{t|t-1}, u_t)\right) \\
    \Sigma_t &= \Sigma_{t|t-1} - G_t C_t \Sigma_{t|t-1}
\end{align}
$R_t$ combines the voltage sensor noise with the conditional variance of the OCV GP at $q_t$.
It is therefore not fixed, but grows wherever the OCV reconstruction is less certain.

At each time step each GP is evaluated at the current charge $q_t$,
where its value is a weighted sum of the basis-point values,
with the weights determined by the kernel.
Because this dependence is linear, these weights form the corresponding entries of the Jacobians $F_t$ and $C_t$,
through which the filter adjusts the basis-point values to bring the predicted voltage closer to the measurement.
Each correction step therefore reshapes $\mathrm{OCV}(q)$ and $R_1(q)$.

Missing data in the current and temperature measurements are filled by linear interpolation,
so $u_t$ is always available.
Gaps in the voltage measurements, by contrast, are handled by the filter itself: it predicts at every step,
propagating the state and its uncertainty forward, but corrects only when a measurement is available.

\subsection*{Filter parametrization and initialization}

Fitting the filter to data requires concrete choices for each GP's kernel and hyperparameters,
the process and observation noise covariances, and the initial conditions for every state.

The joint state combines quantities of different physical scale (volts, amperes, ohms, seconds, kelvin), 
which can leave the filter's covariance matrix poorly conditioned and 
make it hard to choose GP kernel hyperparameters that generalize across cells. 
Before fitting, voltages, currents, and charges are z-score normalized against a fixed nominal range, 
matching the system's operating window: \SIrange{3.3}{4.1}{\volt} (\SI{3.7}{\volt} midpoint) for voltage, 
\SIrange{-50}{50}{\ampere} for current, and \SIrange{-50}{50}{\Ah} for charge. 
For lumped-module models, the voltage normalization is adjusted to the number of series-connected cells ($s=12$).

The recursive GPs for both $\mathrm{OCV}(q)$ and $R_1(q)$ use a squared-exponential kernel 
$k_{SE}(x,x')$ with a constant prior mean.
The prior mean of the OCV GP is set to \SI{3.7}{\volt} and that of the $R_1$ GP to \SI{1}{\milli\ohm}
(\SI{12}{\milli\ohm} for the lumped-module model to reflect the series-connected cells), each an informed initial estimate.
Each GP is represented by 21 basis points, equally spaced over the charge range observed in that cell's fit.
Since voltage, current, and charge are normalized to comparable scales, the GP kernel hyperparameters default to 
$\ell_\text{OCV}=0.5$, $\sigma_\text{OCV}=0.5$, $\ell_{R_1}=0.5$ and $\sigma_{R_1}=0.5$.
This choice aims for a level of smoothing that avoids both losing real features and capturing noise.
These are then rescaled by that cell's own observed voltage and charge span,
so hyperparameters reflect how much of the normalized range that cell's data actually covers.
For some cells and modules, however, the default hyperparameters do not yield a consistent OCV reconstruction.
For these cases, a population-based procedure selects better-fitting hyperparameters instead, described in Note S2.

Beyond the GPs, the filter's scalar parameters and states also need priors, process noise, and initial conditions.
Table~\ref{tab:ekf-tuning} provides an overview of the cell model parameters.
For the lumped-module model, quantities related to voltage or resistance are scaled to reflect
the number of series-connected cells.
The sensor's voltage noise $v_\sigma$ sets how much the filter trusts each voltage measurement.
Together with the OCV GP's uncertainty, it forms $R_t$.
Parameters $R_0$, $\tau$, and $\kappa$ each start from an informed initial value, with an uncertainty set
by how much that parameter could plausibly vary from this guess.
They are treated as static over the measurement window, so their process noise is set to zero.
$v_{\text{rc}}$ is assumed to start at rest, with an initial uncertainty and a small process noise that tolerates 
mismatch from both the zero-order-hold approximation and an initially imprecise parametrization.

The filter is initialized with $q_0 = \SI{0}{\Ah}$ and zero uncertainty, reflecting the known charge state
at the start of the measurement window. 
The process noise for $q$ is also set to zero in this first parametrization run, 
so the filter follows Coulomb counting alone. 
While in this first run $q$ is computed directly from measured current rather than estimated, 
it is still useful to keep it in the state vector instead of treating it as an external input.
When merging additional measurement windows, the initial charge is no longer known with certainty, 
so a nonzero initial uncertainty is needed to let the filter correct the initial guess.
Similarly, for longer runs where Coulomb-counting error can accumulate,
a nonzero process noise lets the filter correct for it.
Correcting $q$ from voltage only becomes reliable once the model, in particular the OCV curve, is characterized. 
Without that, as in this first run, the filter would ``correct'' $q$ toward a wrong estimate
that would then degrade the parameter estimation itself.

\begin{table}[h]
\centering
\caption{EKF tuning parameters for the cell model. Starred (*) values are scaled by the number of series-connected cells for the lumped-module model ($s=12$).}
\label{tab:ekf-tuning}
\begin{tabular}{lccc}
\toprule
State & Initial value & Initial uncertainty ($\sigma_0$) & \shortstack{Process/observation \\ noise ($\sigma_1$, per step)} \\
\midrule
Charge $q$                       & \SI{0}{\Ah}            & \SI{0}{\Ah}              & \SI{0}{\Ah} \\
RC voltage $v_{\text{rc}}$*      & \SI{0}{\volt}          & \SI{0.1}{\milli\volt}    & \SI{0.05}{\milli\volt} \\
Resistance $R_0$*                & \SI{1}{\milli\ohm}     & \SI{0.5}{\milli\ohm}     & 0 (static) \\
Time constant $\tau$             & \SI{800}{\second}      & \SI{5}{\second}          & 0 (static) \\
Arrhenius coefficient $\kappa$   & \SI{2000}{\kelvin}     & \SI{100}{\kelvin}        & 0 (static) \\
Sensor voltage noise $v_\sigma$* & --                     & --                       & \SI{3}{\milli\volt} (measurement) \\
\bottomrule
\end{tabular}
\end{table}

\subsection*{OCV curve alignment and SOH estimation}

Cell capacity and SOH are not parameters in our model.
Instead, the OCV curve of each cell and module is reconstructed based on its own cumulative charge $q$, 
with an unknown initial SOC and no shared reference between them.
Capacity could be read directly off a reconstructed curve
if it spanned the full cell voltage window, which defines \SI{0}{\percent} and \SI{100}{\percent} SOC.
But since the curves cover only a partial voltage and SOC window,
we recover the capacity and initial SOC of every cell (or module) 
by aligning its reconstructed OCV curve against the rest of the population.
Because the curves overlap, cells inform one another,
and every capacity estimate benefits from the voltage range the population covers as a whole.

To align a pair of curves, one is scaled and shifted so that it agrees with the other wherever their voltages overlap.
The OCV GP reconstructs each curve as voltage over charge, $v(q)$. 
Before alignment, each curve is inverted into charge over voltage, $q(v)$.
Each pair of cells (or modules) is compared at 50 evenly spaced points across their shared voltage range:
\begin{equation}
    \alpha_i\,q_i(v) + \beta_i \approx \alpha_j\,q_j(v) + \beta_j,
\end{equation}
where $\alpha_i$ and $\beta_i$ rescale and shift the charge axis of cell $i$.
Because $q_i(v)$ and $q_j(v)$ are known, the equation is linear in the four alignment parameters of the pair.
In the original $v(q)$ frame, they would enter through the curve shapes, making the equation nonlinear.

Collecting every pairwise comparison as rows results in a single linear system,
\begin{equation}
    A\theta \approx b, \qquad \theta = (\alpha_1, \beta_1, \ldots, \alpha_N, \beta_N)^\top,
\end{equation}
that can be solved by ordinary least squares.
The vector $\theta$ collects the alignment parameters $\alpha_i$ and $\beta_i$ of all $N$ cells (or modules).
For $N = 324$ cells, this gives $2N = 648$ unknowns constrained by $N(N-1)/2 = 52\,326$ pairwise comparisons.
Each row of $A$ holds the coefficients of one comparison,
and the corresponding entry of $b$ its right-hand side, which is zero for every pairwise comparison.
Since the comparisons constrain only the relative agreement between cells,
two anchor rows per cell are added to ensure a unique solution,
\begin{equation}
    \alpha_i\,q_i(v_\text{lo}) + \beta_i = 0,
    \qquad \alpha_i\,q_i(v_\text{hi}) + \beta_i = 1
\end{equation}
where $v_\text{lo}$ and $v_\text{hi}$ are the lowest and highest voltages shared by all cells. 
Each row of $A$ has at most four nonzero entries, so the system is solved via the normal
equations $A^\top A\, \theta = A^\top b$, accumulated incrementally as each row is processed
to avoid storing $A$ densely.

The alignment gives the capacity and initial SOC of each cell (or module) over the anchored voltage window.
For cell $i$, $1/\alpha_i$ is the charge stored between $v_\text{lo}$ and $v_\text{hi}$,
and $\beta_i$ its SOC at the start of the measurement window,
with 0 at $v_\text{lo}$ and 1 at $v_\text{hi}$.
To recover the cell's capacity and SOC over the full cell voltage window,
we need an SOC estimate at $v_\text{lo}$ and $v_\text{hi}$:
\begin{equation}
    Q_i = \frac{1}{\alpha_i\,\Delta\mathrm{SOC}}, \qquad
    \mathrm{SOC}_{0,i} = \mathrm{SOC}_\text{lo} + \Delta\mathrm{SOC}\,\beta_i
\end{equation}
where $\Delta\mathrm{SOC}$ is the SOC span between $v_\text{lo}$ and $v_\text{hi}$, and $\mathrm{SOC}_\text{lo}$ the SOC at $v_\text{lo}$.
Both are taken from the OCV curve measured at low power (subsection Validation against reference measurements),
which reaches lower voltages than any curve reconstructed from field data
and is extended linearly to the full cell voltage window,
assuming the curve remains linear beyond its measured range.
The same $\Delta\mathrm{SOC}$ and $\mathrm{SOC}_\text{lo}$ are applied to every cell,
which fixes the absolute level of the reported capacities
and leaves the differences between them to the alignment alone.
Cell and module SOH is obtained by normalizing $Q_i$ by the beginning-of-life capacity given in the datasheet,
$Q_\text{nom} = \SI{100}{\Ah}$ per logical cell.

The module SOH is also determined by aggregating the capacities of its cells,
which accounts for the spread in both capacity and SOC.
Since charge and discharge each stop at the first cell to reach a voltage limit,
the usable module capacity is set by the smallest chargeable and dischargeable capacity
among its cells, $Q_\text{mod}^\text{ch}$ and $Q_\text{mod}^\text{dc}$.
Together they define the module SOH \cite{li_interval_2017,ruther_battery_2025},
\begin{equation}
    \mathrm{SOH}_\text{mod} = \frac{Q_\text{mod}^\text{ch} + Q_\text{mod}^\text{dc}}{Q_\text{nom}}
\end{equation}
with
\begin{equation}
    Q_\text{mod}^\text{ch}  = \min_i \left((1 - \mathrm{SOC}_{i})\,Q_i\right), \qquad
    Q_\text{mod}^\text{dc}  = \min_i \left(\mathrm{SOC}_{i}\,Q_i\right)
\end{equation}
where $i$ runs over the module's twelve cells in series.
Perfect SOC balancing within a module regains the capacity lost to imbalance,
leaving the weakest cell as the only limit,
$Q_\text{mod} = \min_i (Q_i)$.

The aligned curves, averaged at each voltage, form a single composite OCV curve
that serves as reference for the population.
Because the curves cover different, partly overlapping segments,
the composite covers a wider voltage range than any individual cell.
The fitted $\alpha_i$ and $\beta_i$ set the capacity and the initial SOC,
so any error in aligning a curve appears directly in both estimates.
The quality of the alignment therefore indicates how reliable a cell's (or module's) estimates are.
Each curve aligns to the composite only approximately: voltage, current, and temperature
measurement errors perturb the reconstruction, and degradation changes the shape
of the OCV curve, which the alignment cannot absorb.
Their combined effect appears as the residual between a cell's aligned curve and the composite,
from which we quantify the uncertainty on $Q_i$ and $\mathrm{SOC}_{0,i}$.

Because the residuals vary smoothly with voltage rather than from sample to sample,
the classical least-squares covariance, which assumes independence, does not apply.
We therefore model the residual as a zero-mean correlated Gaussian,
with a covariance built from a squared-exponential kernel.
Its length scale is estimated from the population's residual autocorrelation,
and its amplitude from each residual curve.
Propagating this covariance through the alignment gives the uncertainty on $\alpha_i$ and $\beta_i$,
which the delta method transfers to $Q_i$ and $\mathrm{SOC}_{0,i}$.
The resulting $\pm 1\sigma$ uncertainty intervals are wider for cells that agree less with the composite,
for those whose data cover less of the SOC range,
and for those confined to flat regions of the OCV curve.

\subsection*{Simulation and SOC estimation}

Once fitted, the model can be used to simulate arbitrary operating conditions,
driven by the current and temperature inputs alone.
The parameters are frozen at the values reached at the end of the fit,
and the model propagates $q$ and $v_\text{rc}$ through its dynamics
while predicting the terminal voltage, in open loop.
Because the OCV and $R_1$ curves are recovered as functions of charge,
the model applies to any load profile within the charge range covered by the fit.
Prediction errors accumulate, since no measurement corrects the state,
making the open-loop run a test of the model rather than of the filter.

For SOC estimation, the correction step is restored while the parameters stay frozen,
so the state reduces to $q$ and $v_\text{rc}$.
What remains is a conventional ECM-based SOC estimator,
with the reconstructed OCV and $R_1$ curves in place of tabulated ones.
The measurement stays nonlinear in $q$, since both curves are evaluated through
their frozen GPs, so the filter is still an EKF.
Where the reconstruction is uncertain, the GP variance inflates the predicted voltage
uncertainty and down-weights the correction, so extrapolation beyond the fitted range
is visible in the estimate rather than silent.
Charge converts to SOC through the composite fit,
\begin{equation}
    \mathrm{SOC}(t) = \mathrm{SOC}_{0,i} + \frac{q(t)}{Q_i}.
\end{equation}
When the initial SOC is unknown, inverting the reconstructed OCV at the initial voltage
provides a starting value for $q$, which the filter refines as new measurements arrive.

The estimator runs at both levels, so a module's SOC can be estimated directly
or aggregated from its cells.
As with capacity, aggregation follows from the cells that limit charge and discharge.
Evaluated at each step from the cell estimates,
the aggregate is the dischargeable share of the usable capacity,
\begin{equation}
    \mathrm{SOC}_\text{mod}(t) =
    \frac{Q_\text{mod}^\text{dc}(t)}{Q_\text{mod}^\text{dc}(t) + Q_\text{mod}^\text{ch}(t)}.
\end{equation}

We assess SOC accuracy by running the estimator over the same window used for the fit.
As in the fit, charge starts at $q_0 = \SI{0}{\Ah}$ and the RC voltage at rest.
Unlike the fit, $q$ no longer follows Coulomb counting alone:
its initial uncertainty and process noise are nonzero, so the voltage correction
can adjust the integrated charge and absorb accumulated current-measurement error.
The RC voltage keeps its process noise but starts from a much wider prior,
since an estimator applied during operation cannot assume it is at rest.
Table~\ref{tab:soc-tuning} lists the EKF tuning for both states.

\begin{table}[h]
    \centering
    \caption{EKF tuning for the SOC estimation run at cell level. Starred (*) values are scaled by the number of series-connected cells for the lumped-module model ($s=12$).}
    \label{tab:soc-tuning}
    \begin{tabular}{lccc}
        \toprule
        State & Initial value & Initial uncertainty ($\sigma_0$) & Process noise ($\sigma_1$, per step) \\
        \midrule
        Charge $q$                  & \SI{0}{\Ah}   & \SI{0.3}{\Ah}          & \SI{0.3}{\milli\Ah} \\
        RC voltage $v_\text{rc}$*   & \SI{0}{\volt} & \SI{25}{\milli\volt}   & \SI{0.05}{\milli\volt} \\
        \bottomrule
    \end{tabular}
\end{table}

\section*{RESOURCE AVAILABILITY}

\subsection*{Lead contact}

Requests for any additional information about this work should be directed to the lead contact, 
Martín Cornejo (\href{mailto:martin.cornejo@tum.de}{martin.cornejo@tum.de}).

\subsection*{Materials availability}

This study did not generate new materials.

\subsection*{Data and code availability}

The latest version of the associated Julia software is available on GitHub 
(\url{https://github.com/martincornejo/BatteryRecursiveGPs.jl}).
The version used in this paper, together with the dataset, is archived on Zenodo 
(\url{https://doi.org/10.5281/zenodo.22281620}).

\section*{ACKNOWLEDGMENTS}

The authors thank Fabian Dauth and Ramon Schindler (STABL Energy GmbH) for their support with the experimental setup.
This work was funded by the Bavarian Transformation and Research Foundation, Germany, as part of the research project KI-M-Bat (reference number AZ1563-22).

\section*{AUTHOR CONTRIBUTIONS}

Conceptualization, M.C. and A.J.; 
Methodology, M.C. and J.V.S.; 
Software, M.C. and J.V.S.;
Formal Analysis, M.C.; 
Investigation, M.C. and J.M.-S.; 
Data Curation, J.M.-S.;
Visualization, M.C.; 
Writing -- Original Draft, M.C.;
Writing -- Review \& Editing, M.C., J.M.-S., J.V.S., and A.J.;
Supervision, A.J.; 
Funding Acquisition, M.C. and A.J.

\section*{DECLARATION OF INTERESTS}

J.M.-S. is an employee of STABL Energy GmbH, which developed and operates the test system studied in this work. 
The remaining authors declare no competing interests.

\section*{DECLARATION OF GENERATIVE AI AND AI-ASSISTED TECHNOLOGIES}

During the preparation of this work, the authors used Claude (Anthropic) to assist with drafting and to improve the language and readability of the manuscript text.
After using this tool, the authors reviewed and edited the content as needed and take full responsibility for the content of the publication.

\section*{SUPPLEMENTAL INFORMATION}

Document S1: Figures S1--S9 and Notes S1 and S2.


\nocite{rasmussen_gaussian_2006}

\bibliography{references}

\clearpage
\setcounter{figure}{0}
\setcounter{table}{0}
\setcounter{equation}{0}
\renewcommand{\thefigure}{S\arabic{figure}}
\renewcommand{\thetable}{S\arabic{table}}
\renewcommand{\theequation}{S\arabic{equation}}

\begin{center}
  {\LARGE\bfseries Supplementary Material}
\end{center}

\section{Supplementary Figures}

\begin{figure}[htbp]
    \centering
    \includegraphics[width=\textwidth]{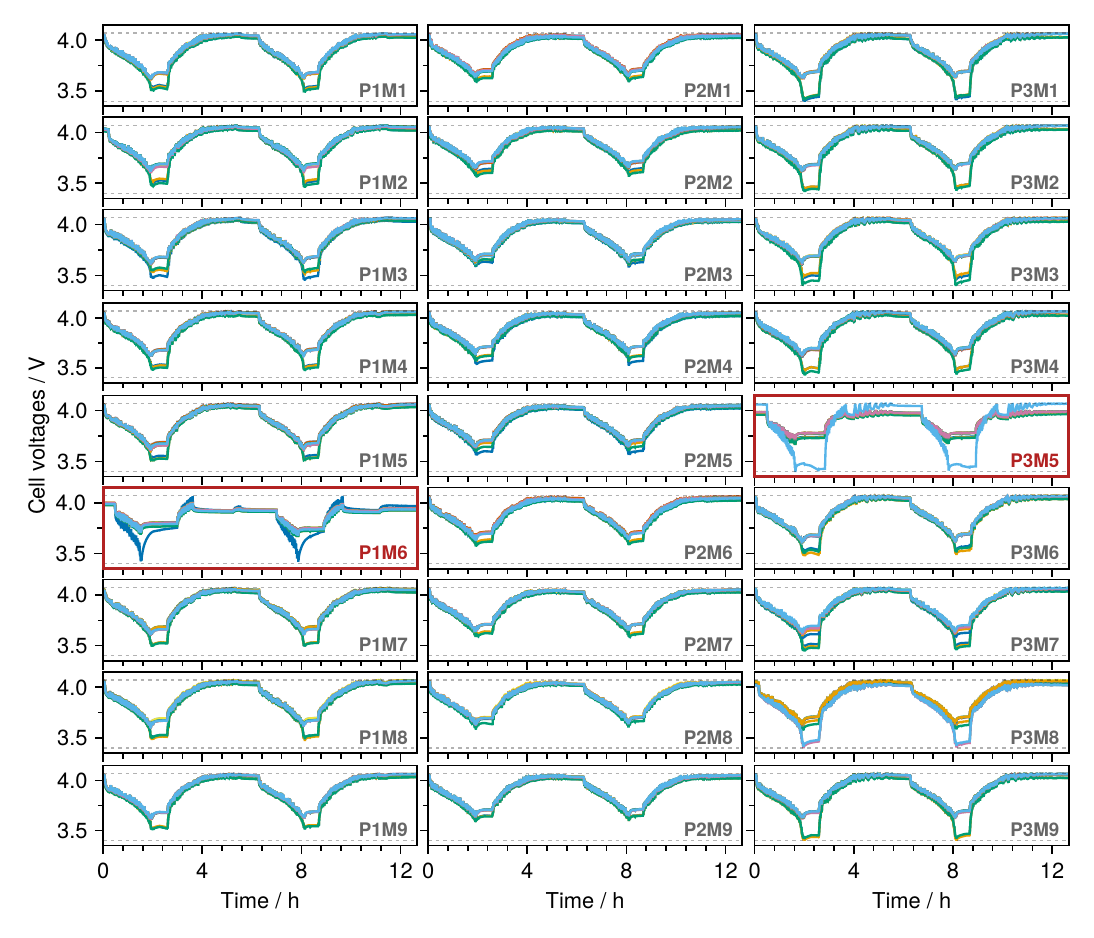}
    \caption{Cell voltages of the full system over the measurement window.
    Each panel shows the twelve cell voltages of one module; rows are module positions M1--M9
    and columns are phases P1--P3.
    Dashed lines mark the operational voltage limits, \SIrange{3.4}{4.07}{\volt}.
    Within every module the cell voltages spread apart over the window,
    most visibly near the end of discharge.
    Red frames mark modules P1M6 and P3M5, which each contain a cell with divergent behavior.}
    \label{fig:dataset-cell-voltages}
\end{figure}

\begin{figure}[htbp]
    \centering
    \includegraphics[width=\textwidth]{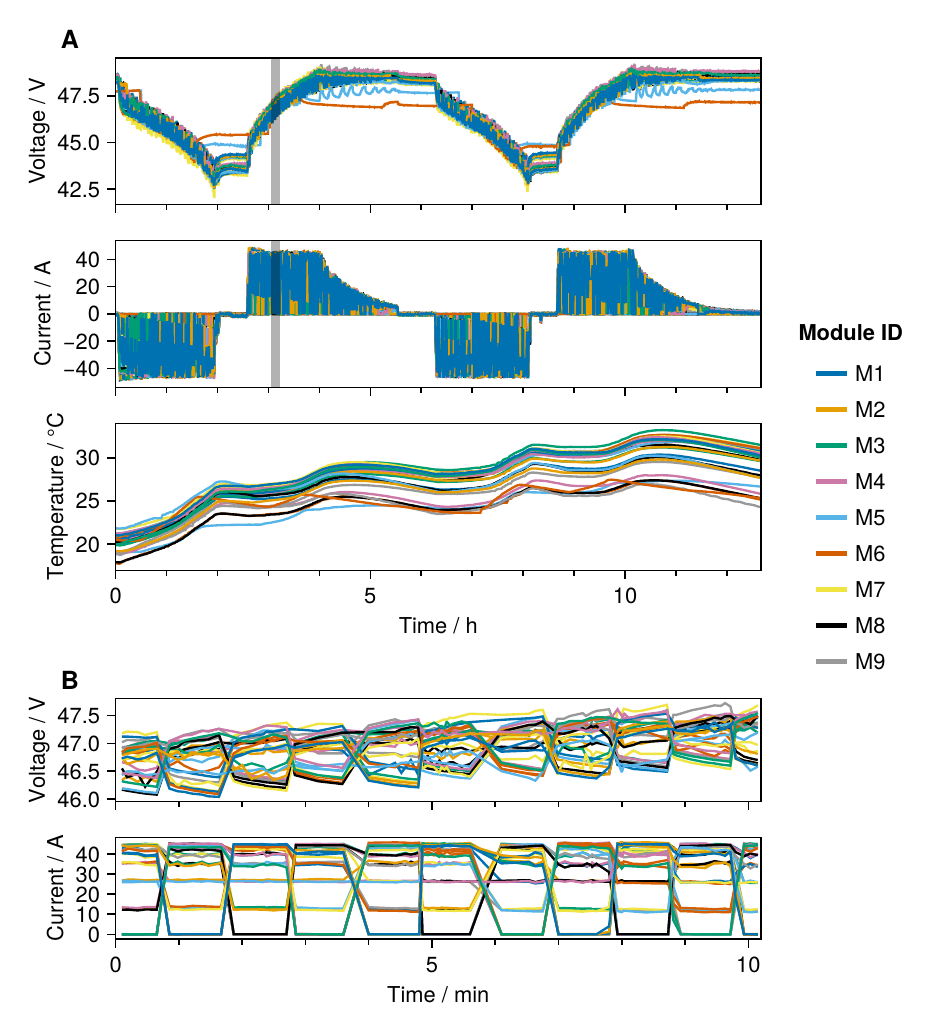}
    \caption{Module voltage, current, and temperature of all 27 modules,
    colored by module position M1--M9 within each phase.
    (A) The full measurement window.
    The system is cycled twice at approximately constant system power,
    but the MMC distributes the phase current unevenly between modules,
    so each module follows its own dynamic load profile.
    Module temperatures differ across the system and drift over the window.
    (B) The shaded \SI{10}{\minute} window in (A).
    The modulation reassigns the load about once a minute, so each module alternates between
    different shares of the phase current, and its voltage steps with every switch.}
    \label{fig:dataset-modules}
\end{figure}

\begin{figure}[htbp]
    \centering
    \includegraphics[width=\textwidth]{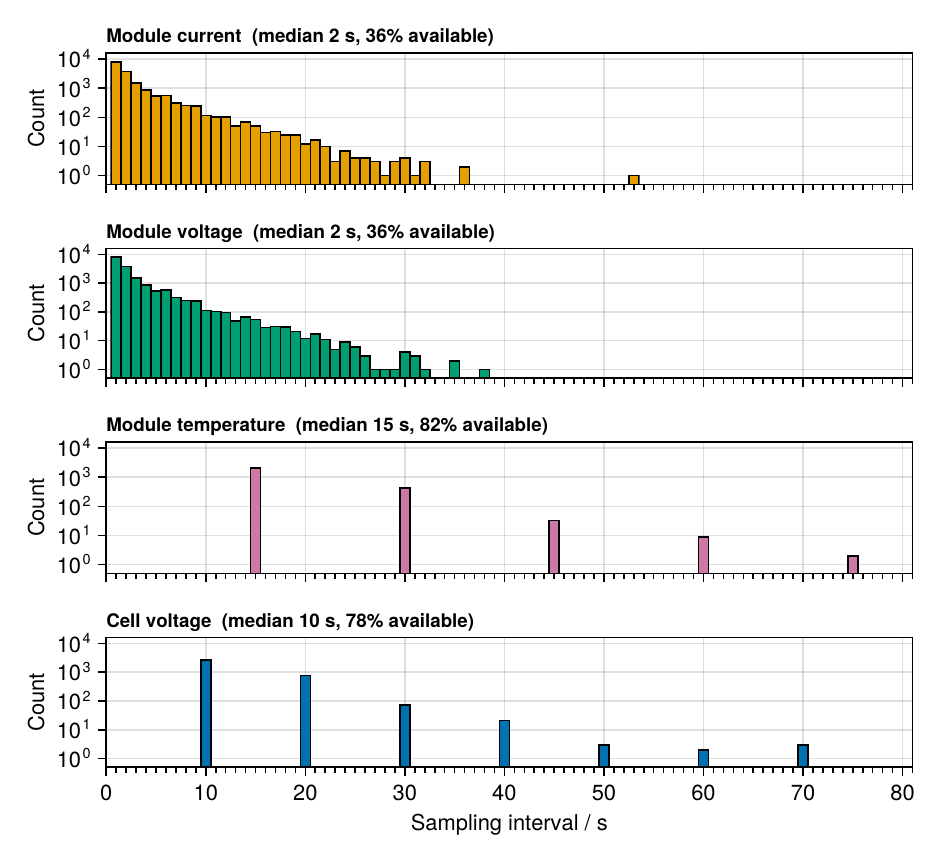}
    \caption{Sampling intervals in the dataset.
    Distribution of the interval between consecutive samples for module current, module voltage,
    module temperature, and cell voltage, on a logarithmic count axis.
    The median interval and the share of available samples are given above each histogram.
    The nominal intervals of \SI{1}{\second} for module current and voltage, \SI{10}{\second} for cell voltage,
    and \SI{15}{\second} for module temperature are not always met.}
    \label{fig:dataset-resolution}
\end{figure}

\begin{figure}[htbp]
    \centering
    \includegraphics[width=0.6\textwidth]{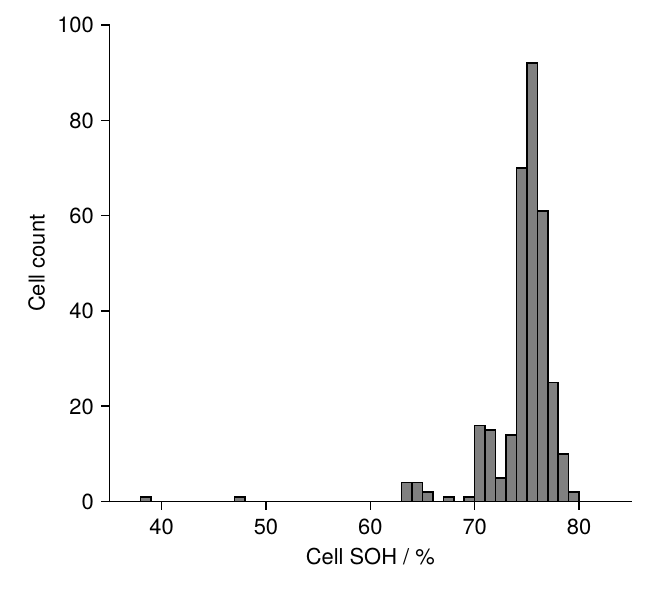}
    \caption{Distribution of estimated cell SOH across the 324 cells.
    Most cells group tightly around a median of \SI{75.4}{\percent}, with a pronounced lower tail.
    Cells P1M6C1 and P3M5C12 are clear outliers,
    at \SI{47.8}{\percent} and \SI{38.0}{\percent}.
    Capacity is obtained by aligning each reconstructed OCV curve against the rest of the population,
    and normalized to the nominal \SI{100}{\Ah} of a logical cell.
    Bin width \SI{1}{\percent}.}
    \label{fig:cell-soh-hist}
\end{figure}

\begin{figure}
    \centering
    \includegraphics{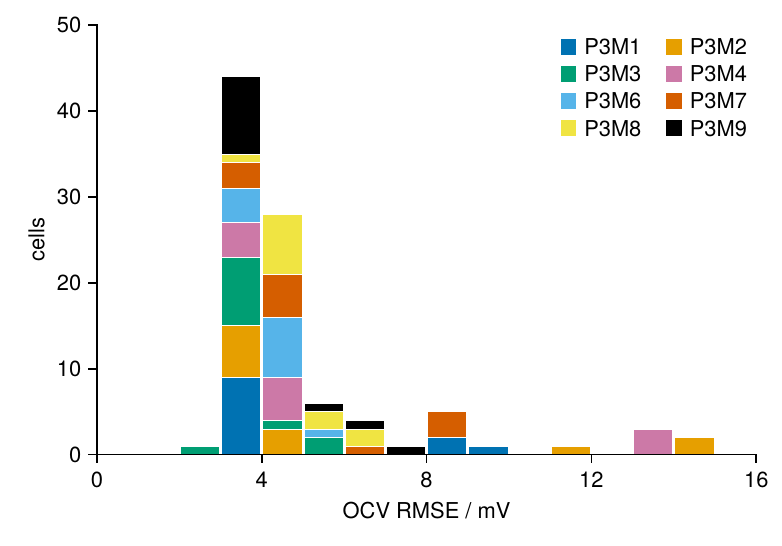}
    \caption{Agreement between the reconstructed and reference pseudo-OCV curves.
    Distribution of the per-cell RMSE over the 96 validated cells,
    stacked by module, with a bin width of \SI{1}{\milli\volt}.
    Most cells lie between \SI{3}{\milli\volt} and \SI{5}{\milli\volt}.
    The tail corresponds to the cells with the lowest SOC,
    whose comparison windows reach the steep knee of the OCV curve.}
    \label{fig:ocv-validation-rmse}
\end{figure}

\begin{figure}[htbp]
    \centering
    \includegraphics[width=0.8\textwidth]{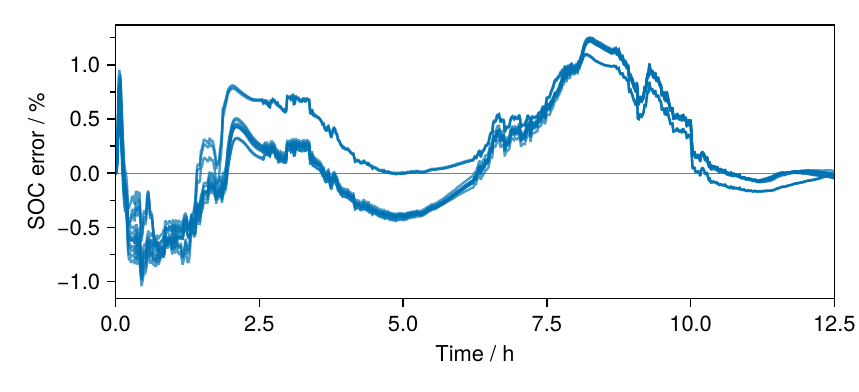}
    \caption{SOC estimation error of the cell-level EKF for the twelve cells of module P1M9, one line per cell,
    relative to Coulomb counting of the external current probe and normalized to each cell's estimated capacity.
    Over the \SI{12.5}{\hour} window the error stays within about \SI{\pm 1.3}{\percent} of each cell's capacity,
    corresponding to less than \SI{1}{\Ah}.}
    \label{fig:ekf-charge-error}
\end{figure}

\begin{figure}[htbp]
    \centering
    \includegraphics[width=\textwidth]{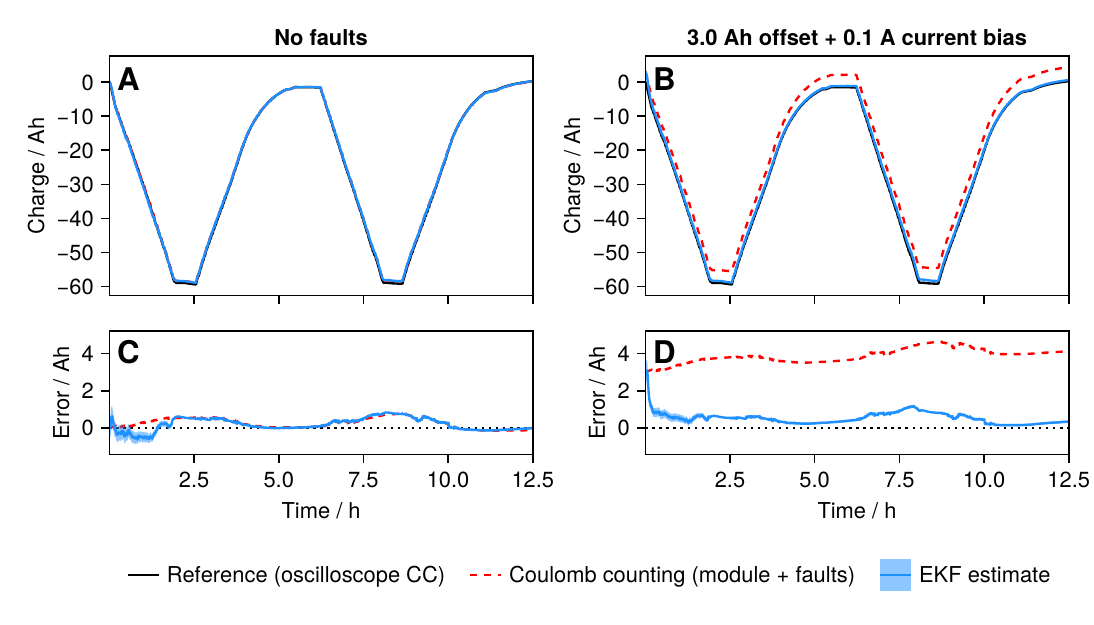}
    \caption{Synthetic fault-injection test of the EKF, illustrated for cell P1M9C2.
    The onboard module current is perturbed by a \SI{3.0}{\Ah} initial offset and a \SI{0.1}{\ampere} constant bias,
    emulating an unknown initial state and a drifting current sensor.
    The perturbed current feeds both Coulomb counting and the EKF.
    (A, B) Cumulative charge from the reference Coulomb count of the external probe (black),
    from Coulomb counting of the onboard current (red dashed),
    and from the EKF estimate with $\pm 2\sigma$ band (blue),
    for the unperturbed (A) and the perturbed (B) current.
    (C, D) Error of both estimates relative to the reference.
    Under perturbation, the Coulomb-count error grows to about \SI{4}{\Ah} by the end of the window,
    whereas the EKF quickly corrects the offset through the voltage measurement
    and then tracks the reference closely.}
    \label{fig:soc-ekf-diagnostic}
\end{figure}

%

\clearpage
\section{Supplemental Notes}

\subsection*{Note S1: Recursive Gaussian process regression}

Gaussian process regression (GPR) has been widely adopted for battery parameter estimation, due to its flexibility in extracting \textit{functional parameters} from data
\cite{schaeffer_health_2024,aitio_learning_2025,zhou_learning_2025}.
Rather than fitting parameters to a predetermined equation,
GPR recovers the shape of a function directly from the data,
constrained only by an assumption of how smoothly it varies.
GPs model functions $y = f(x)$ under the assumption that inputs $x$ and $x'$ that are close produce similar outputs.
GPR determines this correlation from observed data,
then predicts new points by interpolating or extrapolating from the known points.

A GP $f(x)$ starts with a prior, defined by a mean function $m(x)$ (the expected value before any observed data) and a kernel $k(x,x')$ that encodes the correlation structure,
\begin{equation}
    f(x) \sim \mathcal{GP}(m(x), k(x,x'))
\end{equation}
A common choice is the squared exponential (SE) kernel,
\begin{equation}
    k_{SE}(x,x') = \sigma \exp\left(-\frac{(x-x')^2}{2\ell^2}\right)
\end{equation}
where the length scale $\ell$ sets how quickly correlation decays with distance and the variance $\sigma$ controls how much the function is expected to vary around its mean.
Correlation is maximal at $x=x'$ and fades smoothly as $x$ and $x'$ grow apart.
Kernels encoding periodic or non-stationary behavior are also available, and can be combined to reflect more detailed prior knowledge about the function \cite{rasmussen_gaussian_2006}.

For any finite set of inputs $X = \{x_1, x_2, \ldots, x_n\}$, the corresponding outputs $Y = f(X)$ reduce to a multivariate Gaussian distribution
$Y \sim \mathcal{N}(\mu, \Sigma)$ with mean vector $\mu = m(X)$ and covariance matrix $\Sigma = K_{X,X}$,
where the entries are $K_{ij} = k(x_i, x_j)$.
The observed inputs $X$ and new inputs $X^*$ together form another finite set of inputs, so the corresponding outputs $Y$ and $Y^* = f(X^*)$ are jointly Gaussian,
\begin{equation}
    \begin{bmatrix}
        Y \\
        Y^*
    \end{bmatrix}
    \sim \mathcal{N}
    \left(
    \begin{bmatrix}
            m(X) \\
            m(X^*)
    \end{bmatrix},
    \begin{bmatrix}
            K_{X,X}   & K_{X,X^*} \\
            K_{X^*,X} & K_{X^*,X^*}
    \end{bmatrix}
    \right)
\end{equation}
where matrix $K_{X,X^*}$ collects the kernel evaluations between each pair of inputs in $X$ and $X^*$.
Conditioning on the observed values $Y$ then gives predictions $Y^* \mid Y \sim \mathcal{N}(\mu^*,\Sigma^*)$.
Its mean $\mu^*$ gives the predicted values,
and its covariance $\Sigma^*$ gives an uncertainty calibrated to the observations:
\begin{align}
    \mu^* &= K_{X^*,X}K_{X,X}^{-1}(Y-m(X)) + m(X^*)\\
    \Sigma^* &= K_{X^*,X^*} - K_{X^*,X}K_{X,X}^{-1}K_{X,X^*}
\end{align}

A known limitation of GPR is its scalability. Computational cost grows as $\mathcal{O}(n^3)$ in the number of observations $n$, due to the need to invert the $n \times n$ covariance matrix.
Furthermore, since the \textit{training} data are embedded in the GP posterior,
the entire observation history must be stored to make predictions at new points.
This makes GPR prohibitive for large datasets and has motivated approximation methods that reduce the cost \cite{rasmussen_gaussian_2006}.

Huber introduced a recursive formulation of GP regression \cite{huber_recursive_2014} that reframes it as a sequential problem solved with a Kalman filter, avoiding the need to store the observation history.
The latent function is instead approximated by a fixed set of $m$ basis inputs $X_b = \{x_1, \ldots, x_m\}$ and their local estimates $Z_t = f(X_b)$, updated with new observations at each time step $t$.
The state follows a multivariate normal distribution at every time step, $Z_t \sim \mathcal{N}(\mu_t, \Sigma_t)$. Before any observations, $Z_0$ follows directly from the GP prior evaluated at the basis inputs,
$Z_0 \sim \mathcal{N}(m(X_b), K_{X_b,X_b})$.

The latent function evolves as $Z_t = Z_{t-1} + \eta_t$, with process noise $\eta_t \sim \mathcal{N}(0, Q_t)$, giving mean and covariance predictions
\begin{equation}
    \mu_{t|t-1} = \mu_{t-1}, \qquad \Sigma_{t|t-1} = \Sigma_{t-1} + Q_t
\end{equation}
For a static function $Q_t = 0$, the predicted covariance is also unchanged,
while a nonzero $Q_t$ lets the function drift over time.
The observation model relates observation $y_t$ at input $x_t$ to the latent state,
using the GP conditioning recast as a Kalman filter observation equation, where $\epsilon_t \sim \mathcal{N}(0, R_t)$ is the observation noise,
\begin{equation}
    y_t = H_t\big(Z_t - m(X_b)\big) + m(x_t) + \epsilon_t
\end{equation}
$H_t$ is the observation matrix,
\begin{equation}
    H_t = K_{x_t,X_b}K_{X_b,X_b}^{-1},
\end{equation}
and the observation noise covariance $R_t$ combines the sensor noise variance $\sigma_\text{sensor}^2$
with the GP prediction uncertainty at $x_t$, the same residual term as in $\Sigma^*$,
\begin{equation}
    R_t = k(x_t,x_t) - H_tK_{X_b,x_t} + \sigma_\text{sensor}^2
\end{equation}

Because the dynamics and observation model are both linear-Gaussian,
updating $Z_t$ from a new observation $y_t$ follows the Kalman filter correction step,
\begin{align}
    G_t &= \Sigma_{t|t-1} H_t^\top \left(H_t \Sigma_{t|t-1} H_t^\top + R_t\right)^{-1} \\
    \mu_t &= \mu_{t|t-1} + G_t \left(y_t - H_t \mu_{t|t-1} - m(x_t) + H_t m(X_b)\right) \\
    \Sigma_t &= \Sigma_{t|t-1} - G_t H_t \Sigma_{t|t-1}
\end{align}
with the Kalman gain $G_t$.

At any state $Z_t$, the observation model can predict $y^* = f(x^*)$ at an arbitrary $x^*$,
conditioned at $(X_b, Z_t)$ instead of the raw observations,
\begin{align}
    \mu^* &= H^*(\mu_t - m(X_b)) + m(x^*) \\
    \Sigma^* &= H^*\Sigma_t H^{*\top} + R^*
\end{align}
with observation and covariance matrices evaluated at the prediction points
\begin{align}
    H^* &= K_{x^*,X_b}K_{X_b,X_b}^{-1} \\
    R^* &= K_{x^*,x^*} - H^* K_{X_b,x^*}
\end{align}

Because $X_b$ is fixed, $K_{X_b,X_b}^{-1}$ is computed once and reused at every step,
so each update costs only $\mathcal{O}(m^2)$ regardless of how many observations have been processed.
The cost of processing $n$ observations therefore grows as $\mathcal{O}(nm^2)$,
compared to full GP regression's $\mathcal{O}(n^3)$.

\newpage
\subsection*{Note S2: GP hyperparameter selection}

GPs recover functions directly from data, in a non-parametric approach.
Their kernel hyperparameters nonetheless shape that reconstruction, and a poor choice can degrade the fit:
a length scale that is too large smooths away genuine features,
while one that is too short fits measurement noise instead.
Because our battery model estimates $\mathrm{OCV}(q)$, $R_1(q)$, and the scalar ECM parameters jointly,
poorly chosen hyperparameters can let one parameter absorb variation that should be explained by another.

This misattribution is difficult to identify from the voltage signal alone:
over such a limited measurement window, multiple OCV curve shapes, including implausible ones,
can fit the observed voltage about equally well.
The shape of the reconstructed OCV curve can reveal a poor fit directly,
but systematically catching and correcting it requires a reference to check against.
Since no predefined reference is available,
we score each cell's fit by its deviation from a population-wide composite curve.

These reference OCV curves, one for the cell-level fits and one for the module-level fits, are built in two stages.
First, every cell (or module) is fit using the default hyperparameters
(described in the subsection Filter parametrization and initialization of the main text),
and a coarse composite curve is built from the whole ensemble
using the same procedure as described in the subsection OCV curve alignment and SOH estimation of the main text.
Each fit is then scored by its RMSE against this coarse composite,
and cells exceeding a \SI{4}{\milli\volt} threshold are set aside as outliers,
the threshold being \SI{50}{\milli\volt} for modules (twelve series-connected cells).
The composite is then rebuilt from the remaining cells (or modules) only,
giving a cleaner reference unskewed by poor initial fits.

The discarded cells are re-fit over a grid of candidate length scales:
$\ell_{R_1} \in \{0.3, 0.5, 1.0\}$ and $\ell_\text{OCV} \in \{0.5, 0.6, 0.85, 1.0, 1.5, 3, 7, 15\}$.
Each candidate is scored against the reference OCV curve, and the best-fitting one is kept.
The procedure is applied only to the discarded fits, since the composite serves as a plausibility check
rather than a fitting target.
Adapting every cell toward it would suppress the cell-to-cell differences the estimation is meant to resolve.
A small grid search avoids gradient-based optimization,
in which each iteration requires differentiating through the full filter.
Performance varies only gradually with length scale within a reasonable range,
so a coarse search is adequate.

\begin{figure}[h]
\centering
\includegraphics[width=\linewidth]{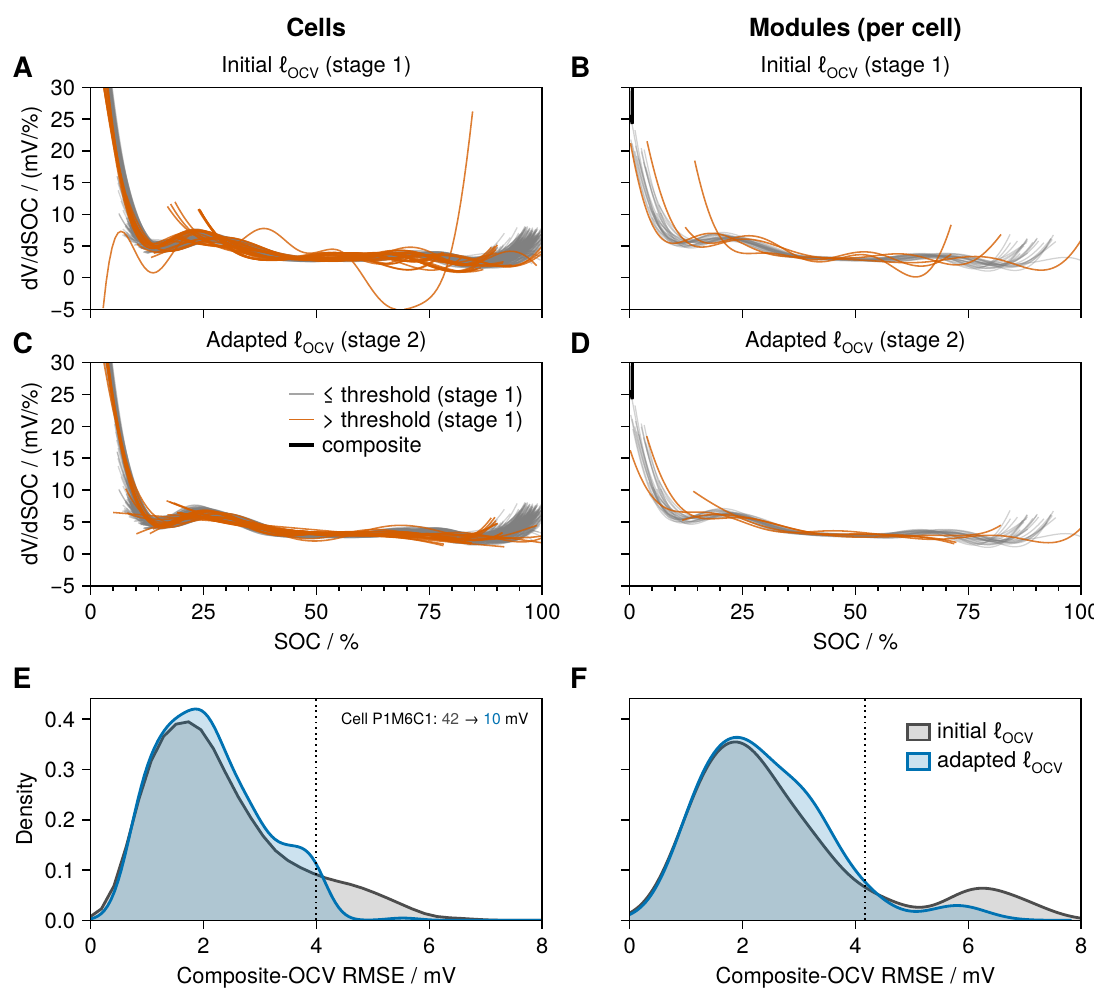}
\caption{Two-stage selection of the GP length scales, for cells (left) and modules (right, per cell equivalent).
(A, B) Reconstructed \dvdsoc~at the default $\ell_\text{OCV}$, and (C, D) after adaptation,
in which $\ell_\text{OCV}$ and $\ell_{R_1}$ are selected jointly.
Cell and module fits that exceed the consistency threshold in the initial fit are drawn in dark orange.
(E, F) Distribution of the RMSE between each reconstructed curve and the composite OCV,
before (gray) and after (blue) adaptation, with the threshold as a dotted line.
Fits beyond the axis range are labeled with their values before and after adaptation.
Adaptation removes most of the implausible curve shapes,
including the curves with negative \dvdsoc~values.}
\label{fig:hyperparam-selection}
\end{figure}

The default hyperparameters yield a consistent OCV reconstruction:
289 of 324 cells and 23 of 27 modules fall within the threshold,
with a median deviation of \SI{1.9}{\milli\volt} from the reference (Figure~\ref{fig:hyperparam-selection}).
For the remaining cells and modules, misfits concentrate near the extremes of the SOC range,
where the reconstructed curve can become implausible, including regions of negative differential voltage
(Figure~\ref{fig:hyperparam-selection}A, B).
After adaptation, 316 of 324 cells and 25 of 27 modules fall within the threshold,
and most curves regain a physically plausible shape (Figure~\ref{fig:hyperparam-selection}C--F).
The worst-performing cell, P1M6C1, improves from \SI{42}{\milli\volt} to \SI{10}{\milli\volt}
only at the largest length scale in the grid ($\ell_\text{OCV}=15$), and still exceeds the threshold.
We interpret this persistent deviation as evidence of an anomalous cell rather than an unresolved fitting problem.

\begin{figure}[h]
\centering
\includegraphics[width=\linewidth]{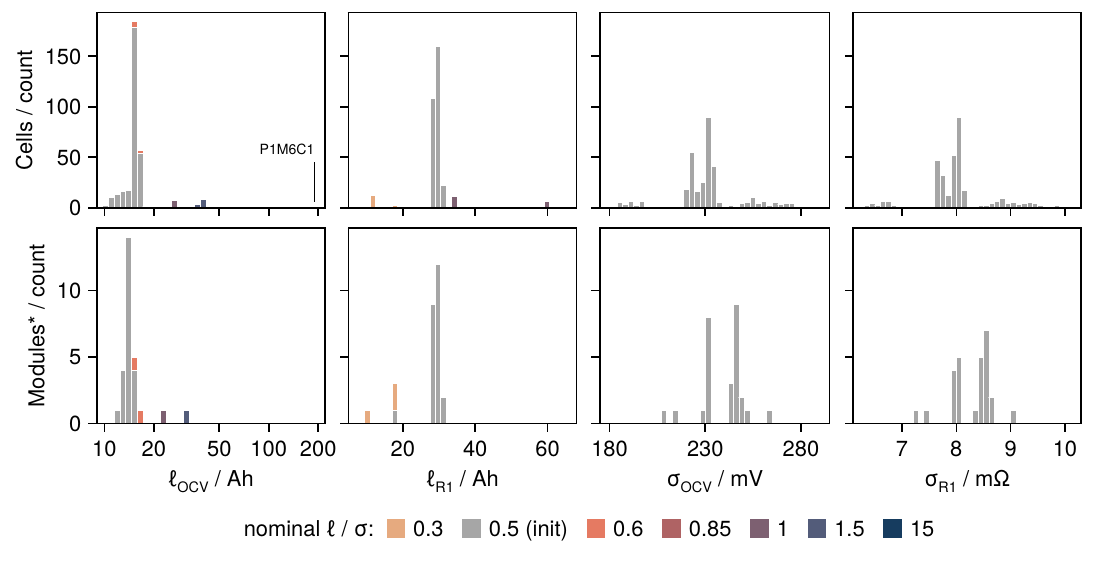}
\caption{GP hyperparameters of all cells (top) and modules (bottom) in physical units.
Columns give the two length scales, $\ell_\text{OCV}$ and $\ell_{R_1}$ in ampere-hours,
and the two prior standard deviations, $\sigma_\text{OCV}$ in millivolts and $\sigma_{R_1}$ in milliohms,
with $\ell_\text{OCV}$ on a logarithmic axis.
Bars are colored by the selected nominal length scale, gray for the default.
Since only the length scales are adapted, the two standard-deviation columns are uniformly gray.
The spread within a color reflects the range each fit covers.
Module standard deviations are expressed per cell equivalent, while the length scales are common to both levels.
On this basis the module and cell hyperparameters coincide.}
\label{fig:hyperparam-scales}
\end{figure}

Figure~\ref{fig:hyperparam-scales} shows the resulting GP hyperparameters in physical units. The module hyperparameters coincide with the cell hyperparameters when expressed per cell equivalent, confirming that the same parametrization serves both model levels.

\end{document}